\documentclass[prd, showpacs, twocolumn,superscriptaddress,nofootinbib]{revtex4}

\usepackage{graphicx}
\usepackage[normalem]{ulem}	
\usepackage[usenames]{color}

\def\lsim{\raise0.3ex\hbox{$\;<$\kern-0.75em\raise-1.1ex
\hbox{$\sim\;$}}}
\def\gsim{\raise0.3ex\hbox{$\;>$\kern-0.75em\raise-1.1ex
\hbox{$\sim\;$}}}

\let\vev\VEV
\def\be{\begin{equation}}
\def\ee{\end{equation}}
\def\ba{\begin{eqnarray}}
\def\ea{\end{eqnarray}}

\begin{document}
\title{
Potential of a Neutrino Detector in the ANDES Underground 
Laboratory for Geophysics and Astrophysics of Neutrinos}
\author{P.~A.~N.~Machado}
\email{accioly@fma.if.usp.br}
\affiliation{
Instituto de F\'{\i}sica, Universidade de S\~ao Paulo, 
 C.\ P.\ 66.318, 05315-970 S\~ao Paulo, Brazil}
\affiliation{Institut de Physique Th\'eorique, CEA-Saclay, 91191 Gif-sur-Yvette, France}
\author{T.~M\"uhlbeier}
\email{muhlbeier@fis.puc-rio.br} \affiliation{Departamento de F\'{\i}sica,
  Pontif{\'\i}cia Universidade Cat{\'o}lica do Rio de Janeiro,
  C. P. 38071, 22452-970, Rio de Janeiro, Brazil}
\author{H.~Nunokawa}
\email{nunokawa@puc-rio.br} \affiliation{Departamento de F\'{\i}sica,
  Pontif{\'\i}cia Universidade Cat{\'o}lica do Rio de Janeiro,
  C. P. 38071, 22452-970, Rio de Janeiro, Brazil}
\author{R.~Zukanovich Funchal} 
\email{zukanov@if.usp.br} \affiliation{
  Instituto de F\'{\i}sica, Universidade de S\~ao Paulo,
  C.\ P.\ 66.318, 05315-970 S\~ao Paulo, Brazil}
\affiliation{Institut de Physique Th\'eorique, CEA-Saclay, 91191 Gif-sur-Yvette, France}


\begin{abstract} 
The construction of the Agua Negra tunnels that will link Argentina 
and Chile under the Andes, the world longest mountain range, 
opens the possibility to build the first deep underground laboratory 
in the Southern Hemisphere. 
This laboratory has the acronym ANDES (Agua Negra Deep Experiment
Site) and its overburden could be as large as $\sim 1.7$ km of rock,
or 4500 mwe, providing an excellent low background environment to
study physics of rare events like the ones induced by neutrinos and/or
dark matter.
In this paper we investigate the physics potential of a few kiloton size
liquid scintillator detector, which could be constructed in the ANDES
laboratory as one of its possible scientific programs. 
In particular, we evaluate the impact of such a detector
for the studies of geoneutrinos and galactic supernova neutrinos 
assuming a fiducial volume of 3 kilotons as a reference size.
We emphasize the complementary roles of such a detector to the 
ones in the Northern
Hemisphere neutrino facilities through some advantages due to
its geographical location.
\end{abstract} 

\maketitle

\section{Introduction}
\label{sec:intro}

After the pioneering neutrino experiments performed at Homestake
(South Dakota, USA)~\cite{Cleveland:1998nv} and Kamioka (Gifu,
Japan)~\cite{Hirata:1987hu}, and the great achievements by
Super-Kamiokande~\cite{sk-atm}, KamLAND~\cite{KamLAND}, both in
Kamioka, and SNO~\cite{Ahmad:2002jz} (Sudbury, Canada) experiments
which provided strong evidence of neutrino oscillation, it has been well
recognized that deep underground laboratories can offer an excellent
environment for neutrino experiments as well as for a variety of
interesting scientific programs which include several different
fields, from particle physics, astrophysics, nuclear physics, to
biology, geology and geophysics. See \cite{Smith-review-taup11} for a
review of the world's underground laboratories.

Experiments searching for very rare events, such as the ones induced
by neutrinos or dark matter interactions, proton decay or performing
low energy nuclear cross section measurements, cannot be carried out
at the Earth's surface mainly due to the backgrounds induced by cosmic
rays. For these experiments a reduction of the cosmogenic backgrounds is
crucial. This can be accomplished by having enough rock overburden,
making clear the reason for going deep underground.

Recently it was proposed~\cite{Andes-Website} to build the first
underground laboratory in the Southern Hemisphere by digging a cave
off one of the two 14 km long Agua Negra tunnels.  These tunnels will
be constructed under the Andes, the longest continental mountain range
in the world, to connect Chile's region IV and Argentina's San Juan
province. They will provide a link between the port of Coquimbo,
Chile (on the Pacific Ocean), and the port of Porto Alegre,
Brazil (on the Atlantic Ocean), and several nearby cities, in
order to facilitate trade between Asia and MERCOSUR (Mercado Com\'un
del Sur or Common Southern Market in English) which is an economic and
political agreement among Argentina, Brazil, Paraguay and Uruguay.
While the exact location of the laboratory inside the tunnel is still
under study, the rock overburden could be as large as $\sim$ 1.7 km,
allowing to significantly reduce backgrounds from cosmic ray origin.
The name given to the proposed laboratory is ANDES (Agua Negra Deep
Experiment Site).

If such an underground laboratory is constructed, it could provide a 
variety of interesting scientific opportunities  
for dedicated studies of neutrinos, dark matter searches and  
nuclear astrophysics~\cite{Andes-Website}, among other things. 
The current preliminary design of the ANDES laboratory is as
follows~\cite{Bertou-workshop3}.  There will be two large experimental
halls with dimensions of $21 \times 23\times50$ m$^3$ and 
$16\times 14 \times 40$ m$^3$, and one smaller hall for offices and
multidisciplinary experiments with size of $17\times15\times 25$ m$^3$.
In addition, there will be two experimental pits, one is a smaller
ultra-low radiation pit with 8 m of diameter and 9 m of height, the
other is a large single experiment pit with 30 m of diameter and 30 m
of height.
We are particularly interested in this large pit where a liquid
scintillator detector could be installed and used in a possible
neutrino program for the ANDES.

As demonstrated by KamLAND~\cite{KamLAND},
Borexino~\cite{Borexino-solar,Borexino-geo}, as well as the recent
$\theta_{13}$ reactor experiments, Double Chooz~\cite{Abe:2011fz},
Daya Bay~\cite{An:2012eh} and RENO~\cite{collaboration:2012nd}, liquid
scintillator detectors have a very good capability to observe
$\bar{\nu}_e$ through the inverse beta decay reaction, $\bar{\nu}_e +
p\to n + e^{+}$. They also can work with a low energy threshold  
and provide good energy resolution.
As a possible candidate for the ANDES neutrino detector one could
consider a KamLAND~\cite{KamLAND}, Borexino~\cite{Alimonti:2000xc} or
SNO+~\cite{Kraus:2010zzb} like liquid scintillator detector with a
fiducial mass of a few kt.
In this paper, we assume a liquid scintillator detector based on alkyl
benzene (C$_6$H$_5$C$_{12}$H$_{25}$), which will be used for the SNO+
detector, with a fiducial mass of 3 kt, containing $\sim 2.2 \times
10^{32}$ free protons as targets, as our reference neutrino detector
at the ANDES, unless otherwise stated.

We focus on the detection of neutrinos originating from some
radioactive elements inside the Earth, the so called {\it geoneutrinos},
and neutrinos coming from a core collapse supernova (SN) in our galaxy
(hereafter SN implies core collapse supernova).  For reviews on these
subjects, see, for example,
Refs.~\cite{Fiorentini:2007te,Enomoto-thesis-2005} for geoneutrinos
and Ref.~\cite{Raffelt:2010zza} for SN neutrinos.  Some preliminary
results of this work can be found in
Refs.~\cite{Andes-talk1,Andes-talk2,Andes-talk3}.

The first successful observations of geoneutrinos by
KamLAND~\cite{Araki:2005qa} and Borexino~\cite{Bellini:2010hy} open a
new window to study the chemical composition of the Earth.
It is believed that the geoneutrino flux has a local
dependency~\cite{Fiorentini:2007te}, hence having more detectors in
different parts of the Earth capable of measuring such neutrinos is
highly welcome.
Since the ANDES laboratory is surrounded by a thick continental crust,
we expect a geoneutrino flux larger than at Kamioka or 
Gran Sasso, which is interesting to confirm experimentally.  Moreover,
compared to other existing underground laboratories, there are
very few nuclear reactors around the ANDES location - a valuable
  advantage as reactor neutrinos are one of the main backgrounds for
  the measurement of geoneutrinos.

After the historical observations of SN neutrinos from SN1987A
occurred in the Large Magellanic Cloud by
Kamiokande~\cite{Hirata:1987hu}, IMB~\cite{Bionta:1987qt}, and
Baksan~\cite{Alekseev:1988gp} detectors, it is well understood that
such SN neutrinos can play a very important role in uncovering the
physics of supernovae as well as some properties of the neutrinos
themselves, such as mass, lifetime, magnetic moment,
etc~\cite{Raffelt:1996wa}.
The low rate of nearby SN, which occurs close enough to the Earth so
that it can be observed also by neutrinos, is another strong reason
for having as many simultaneously operating neutrino detectors as
possible, in order to take advantage of such a rare opportunity.
The additional new neutrino detector would also help in making a quick
alert to astronomers about the occurrence of a nearby SN event through
the SuperNova Early Warning System (SNEWS)
network~\cite{Antonioli:2004zb}.

The ANDES neutrino detector, if constructed, can certainly make some
relevant contribution to SN neutrino observations.  Furthermore, the
location of the ANDES laboratory in the Southern Hemisphere can
provide a better chance to observe the Earth matter effect for SN
neutrinos by combining with other detectors in the Northern
Hemisphere.
If Earth matter effect is observed, the neutrino mass hierarchy
may be determined and, at the same time, evidence that different SN 
neutrino flavors manifest significantly different energy spectra can be 
provided.

The organization of this paper is as follows.  In
Sec. \ref{sec:geoneutrinos} we discuss in detail the potential of the
ANDES neutrino detector for geoneutrino observations.
In Sec.~\ref{sec:SN-neutrinos} we discuss SN neutrino observation 
at the ANDES neutrino detector. 
Sec.~\ref{sec:conclusions} is devoted to discussions and 
conclusions of our results. 
While we follow previous works for most of the calculations done in
this work, for the sake of completeness and to be self-contained,  we 
describe some details of our numerical calculations in the Appendices.

\section{Geoneutrinos}
\label{sec:geoneutrinos}

\subsection{Introduction}
\label{subsec:geo-intro}

The deep interior of the Earth, governed by high pressure and temperature, 
is the last frontier on our planet, which has not yet been 
explored by a human being. 
The deepest hole ever made so far is 12.3 km down
from the Earth's surface~\cite{Fiorentini:2007te}, 
only about 0.1\% of the Earth's diameter, 
and so even the top of the mantle has not yet been reached. 
In the near future, a direct access to the upper mantle will be
possible as part of the missions of the international scientific
research program named IODP (Integrated Ocean Drilling
Program)\cite{IODP}.  However, to reach the lower part of the Earth's
mantle, located at a depth of about 660 km from the surface, seems to be an
impossible task.

So far,
there are basically two different approaches to overcome the direct
inaccessibility of the Earth's underground below 10 km: seismology and
geochemistry. Seismological data permit us to indirectly reconstruct
the matter density profile of the whole Earth. Geochemistry, however,
can only access the composition of the Earth close to the surface.
For that one uses various rock samples coming from the Earth's crust
as well as very limited samples from the top of the mantle, thanks to
volcanic activity and orogeny.

In 2005 the KamLAND experiment~\cite{Araki:2005qa} reported for the
first time the detection of $\bar{\nu}_e$ coming from the decay chains
of the radioactive isotopes $^{238}$U and $^{232}$Th in the Earth, by
using the inverse beta decay reaction $\bar{\nu}_e + p \to n + e^+$.
These geoneutrinos can provide useful information not only relevant to
the Earth's interior chemical composition but also shed light on the
source of the terrestrial heat production, opening a window to a new
scientific field, {\it neutrino geophysics} or {\it neutrino
  geoscience}.
In 2010, another experiment, Borexino, located in the Gran Sasso
Laboratory in Italy also reported the measurement of
geoneutrinos~\cite{Bellini:2010hy}, further contributing to the start
of neutrino geoscience.  A partial list of previous works on
geoneutrinos can be found in Refs.~\cite{Krauss:1983zn,Raghavan:1997gw,
  geo-nu-fiorentini,geo-nu-bari,Nunokawa:2003dd,Fiorentini:2007te}.

\begin{figure*}[!t]
\begin{center}
\vglue -2.6cm
\hglue 0.3cm
\includegraphics[width=0.95\textwidth]{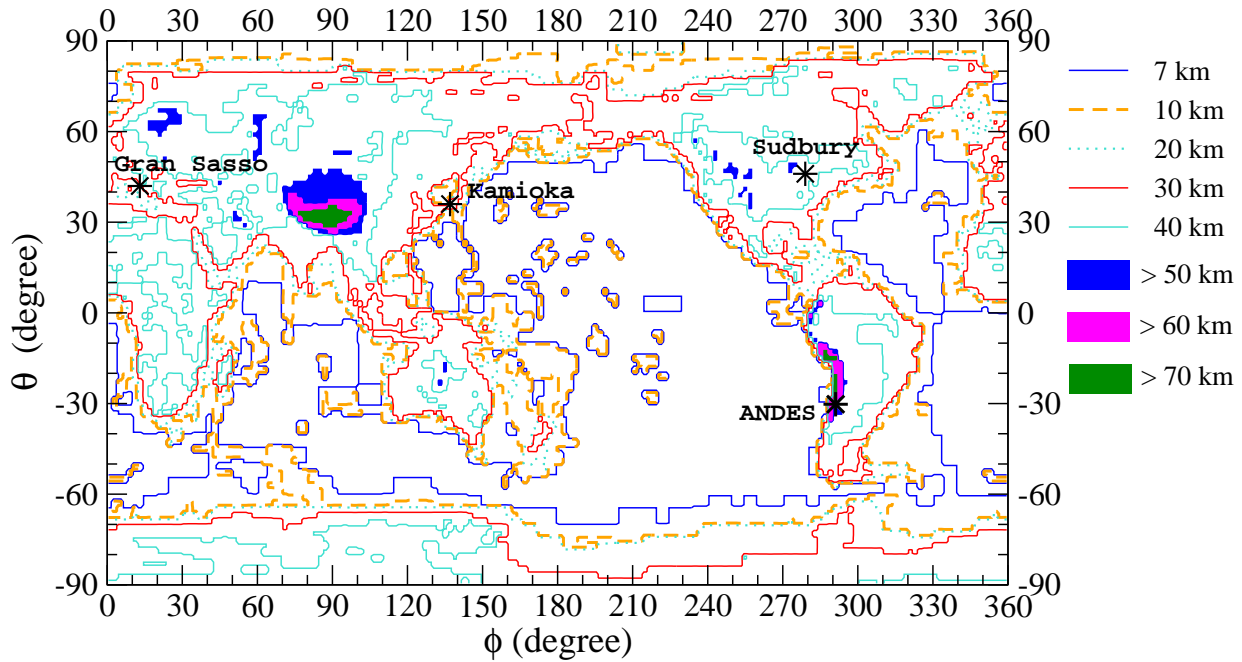}
\end{center}                          
\vglue -2.2cm
\caption{Isocontour map of the Earth's crust thickness (in km) in the
  plane of $\theta$ (latitude)-$\phi$ (longitude) based on the model
  found in Ref.~\cite{geoneutrinos-gabi-etal}.  The positions of the
  Gran Sasso, Kamioka, Sudbury and ANDES underground laboratories are
  also indicated by asterisks.  }
\label{fig:thickness-crust}
\end{figure*}

It is estimated that the entire Earth generates about 40 TW,
corresponding to $\sim$ 10,000 reactors. 
It has been considered that most (or all) 
of the heat is generated 
by the energy deposited by the decay of radioactive elements 
like U, Th and K in the Earth's interior. 
By measuring geoneutrinos, which are the direct product of such decays,
one can infer the total amount of U and Th inside the Earth.  We note
that the amount of K can not be inferred directly by the current
detection method since the energy of geoneutrinos coming from K is
below the threshold of the inverse beta decay reaction.

The measured geoneutrino flux, reported recently by the KamLAND
experiment~\cite{KamLAND-nature-geo2011}, is $4.3^{+1.2}_{-1.1} \times
10^6$ cm$^{-2}$s$^{-1}$. By taking into account neutrino oscillations, this 
corresponds to a total emitted flux of $7.4^{+2.1}_{-1.9} \times
10^6$ cm$^{-2}$s$^{-1}$.
On the other hand, Borexino experiment~\cite{Borexino-geo} reported in
terms of the observed number of events,
$3.9^{+1.6}_{-1.3}(^{+5.8}_{-3.2})$ events/(100 ton yr) at 68\%
(99.73 \%) CL.  This result is about 70 \% higher than that obtained
by KamLAND~\cite{KamLAND-nature-geo2011}, though both results are
consistent with each other within the experimental and theoretical
uncertainties.

By combining the results from KamLAND and Borexino, the observed geoneutrino
flux corresponds to a heat production of $20^{+8.8}_{-8.6}$
TW~\cite{KamLAND-nature-geo2011}.
While Borexino seems to favor, KamLAND results disfavor the so called
fully radiogenic model, the model where all the terrestrial heat comes
from the decay of the radioactive elements in the Earth crust and
mantle. KamLAND alone disfavors this model at 98.1\% CL while the 
combined data of these two experiments slightly reduces the
significance of this rejection to 97.2\%
CL~\cite{KamLAND-nature-geo2011}.

\subsection{Why measure geoneutrinos in the ANDES?}
\label{subsec:why-andes}

For various reasons it would be very interesting if the measurement of
geoneutrinos can also be done at the ANDES Laboratory.  First we note
that the location of the ANDES laboratory, 30$^\circ$15' S and
69$^\circ$53' W, is surrounded by the Andes mountain range which means
that the thickness of the crust around the laboratory is significantly
larger than the average Earth crustal thickness, leading to an
expected larger geoneutrino flux.  This is because in the Earth's
crust the concentration of U and Th is expected to be significantly
larger than in the deeper mantle.

\begin{figure}[!h]
\begin{center}
\vglue -0.6cm
\hglue -0.2cm
\includegraphics[width=0.71\textwidth]{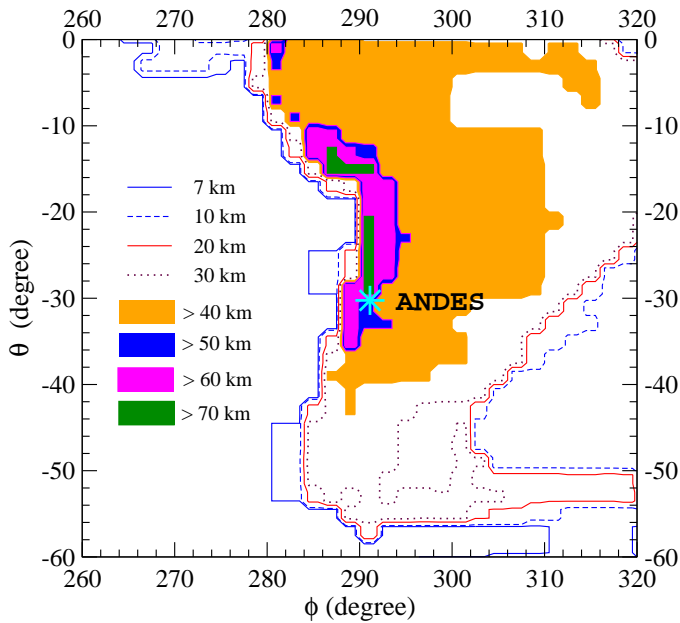}
\end{center}                          
\vglue -1.8cm
\caption{Same as in Fig.~\ref{fig:thickness-crust} but
around the ANDES laboratory. }
\label{fig:thickness-crust-andes}
\end{figure}

In Fig.~\ref{fig:thickness-crust} we present the 
isocontour map of the Earth crust thickness based 
on the model found in Ref.~\cite{geoneutrinos-gabi-etal}.
In Fig. \ref{fig:thickness-crust-andes} we present 
the magnified version of Fig.~\ref{fig:thickness-crust} 
around the ANDES laboratory. 
From these figures, we can appreciate the difference of  
the local crust thickness around the ANDES laboratory in comparison 
with other locations on our planet.
Indeed, roughly speaking, the expected geoneutrino flux at the ANDES
laboratory site is larger than that for KamLAND and Borexino by about
30 \% and 20 \%, respectively, as we will see below. 
It is important to confirm such local dependence of the geoneutrino fluxes.
Because of such a local (site) dependence, it would be also very
interesting to measure the geoneutrino flux at a location surrounded
by oceanic crust, such as in Hawaii~\cite{Hanohano}, where the main
contribution to the geoneutrino flux is expected to come from the decay of 
radioactive elements in the mantle.

Second, there are very few nuclear reactors near the ANDES laboratory
which is a great advantage for measuring geoneutrinos.  As it is well
know from the results of the KamLAND experiment, low energy $\bar
\nu_e$ produced by nuclear reactors is one of the most important
backgrounds for geoneutrino observation.
Indeed, the Borexino detector, despite being smaller than KamLAND,
demonstrated so far a better performance than KamLAND as far as geoneutrino
measurements are concerned.  This seems to be due to the fact that
Borexino, located in the middle of Italy where there are no nearby reactors,
is exposed to a much lower $\bar \nu_e$ background from reactor origin.

Taking into account the nearest Argentinian nuclear reactors, the
Embase 2.1 GW (thermal power) as well as the Atucha I 1.2 GW and
Atucha II 2.1 GW reactors, which are located, respectively, 560 km and
1080 km away from the ANDES Laboratory, we have estimated a total
reactor background of 8.8 events/yr/3 kt and 2.2 events/yr/3 kt in the
geoneutrino energy range. These numbers are given in the absence of
neutrino oscillations, which will further reduce them
if taken into account.  
Though we have
considered only the contribution from these nearby reactors, our
estimation is similar to the one which can be inferred from Fig.~2 of
Ref.~\cite{Lasserre:2010my} which shows the world map of the
isocontours of the expected number of events induced by neutrinos
coming from 201 nuclear reactor power stations all over the world.
The reactor neutrino background we found for ANDES is more than 10
times smaller than the one expected for Borexino, 5.7 events/yr/100
tons (total number of reactor $\bar{\nu}_e$ induced events in the
presence of oscillation)~\cite{Borexino-geo}.

\subsection{Calculating the geoneutrino flux}
\label{subsec:geo-analysis-procedure}

We follow our (two of the authors of this paper) previous
work~\cite{Nunokawa:2003dd} with some update and improvements, 
in order to compute the differential flux of $\bar{\nu}_e$ produced in the
decay chain of radioactive isotopes $^{238}$U and $^{232}$Th that will
be measured at a detector position $\mathbf{r}$ on the Earth, which
can be expressed by the following integral performed over the Earth's 
volume $V_{\oplus}$,
\begin{widetext}
\vglue -0.5cm
\begin{equation}
\label{eq:eqflux}
\displaystyle 
\frac{d \Phi_{\bar{\nu}_e}(\mathbf{r})}{d E_{\bar \nu_e}} 
 = 
\sum_{k=\text{U,Th}}\int _{V_{\oplus}} d^3\mathbf{r}'\frac{ \rho(\mathbf{r}') } 
{4\pi|\mathbf{r}-\mathbf{r}'|^2} \,                                  
\frac{c_k(\mathbf{r}')n_k}{\tau_k m_k}\, 
\times P_{\bar \nu_e}(E_{\bar \nu_e},|\mathbf{r}-\mathbf{r}'|) 
\times f_k(E_{\bar \nu_e}), 
\end{equation}
\end{widetext}
where $\rho (\mathbf{r})$ is the matter density, $c_k(\mathbf{r})$, 
$\tau_k$, $m_k$ and $n_k$ are, respectively, the mass abundance, half-life, 
atomic mass and the number of $\bar \nu_e$ emitted per decay chain
corresponding to element $k=^{238}$U, $^{232}$Th, 
$f_k(E_{\bar \nu_e})$ is the normalized spectral function for 
element $k$~\cite{spectrum}. 

We will assume the concentrations of U and Th, $c_k(\mathbf{r})$, take
different values in the Earth's crust and mantle layers as given in
Table~\ref{tab1}. These are our reference values which are based on
Ref.~\cite{Enomoto-thesis-2005}.  
Oceanic and continental Earth's crust are
divided, respectively, into two and four layers whereas the Earth's
mantle is divided into two layers (see Table~\ref{tab1}).  We assume
no U and Th in the Earth's core.  The so called fully radiogenic model
assumes higher U, Th and K abundances in the mantle in order that the
total observed Earth's heat can be fully explained by the decay of these 
radio active elements.
We have used the Earth crust model taken from 
Ref.~\cite{geoneutrinos-gabi-etal} (shown in Figs.~\ref{fig:thickness-crust} 
and  \ref{fig:thickness-crust-andes}).  

$P_{\bar{\nu}_e}(E_{\bar \nu_e}, |\mathbf{r}-\mathbf{r}'|)$ 
describes the survival probability of $\bar{\nu}_e$ 
produced at $\mathbf{r}'$ but measured at $\mathbf{r}$, 
which can be averaged out, as a good approximation, and
bring out from the integral the term:
\begin{widetext}
\vglue -0.5cm
 \begin{equation}
\langle P_{\bar \nu_e} \rangle  \simeq \left\langle \sin^4
\theta_{13} \right.
\left. +\cos^4 \theta_{13}\left( 1- \sin^22\theta_{12} \sin^2
\left[\frac{\Delta m^2_{21}}{4E_{\bar{\nu}_e}}|\mathbf{r}-\mathbf{r}'|
  \right] \right) \right\rangle 
\simeq \sin^4 \theta_{13} + 
\cos^4\theta_{13}\left(1-\frac{1}{2} \sin^22\theta_{12} \right) \simeq
0.55,
\end{equation}
\end{widetext}
where $\Delta m^2_{21} \equiv m_2^2- m_1^2 \simeq 7.5\times 10^{-5}$
eV$^2$, $m_1$, $m_2$ being the neutrino masses, $\sin^2\theta_{12} =
0.31$ and $\sin^2 \theta_{13} =0.025$~\cite{global-analysis}.
In this work we use the standard neutrino mixing 
parameterization found in Ref.~\cite{Nakamura:2010zzi}.

\begin{table}[t!]
\begin{tabular}{|l|c|c|}
\hline
Layer & $c_{\rm U}$ ($\mu$ g/g) & $c_{\rm Th}$ ($\mu$ g/g) \\
\hline\hline
Oceanic Sediment & 1.68 & 6.91 \\
Oceanic Crust    & 0.1 & 0.22 \\
\hline
Continental Sediment & 2.8 & 10.7 \\
Upper Continental Crust & 2.8 & 10.7 \\
Middle Continental Crust & 1.6 & 6.1 \\
Lower Continental Crust & 0.2 & 1.2 \\ 
\hline
Upper Mantle & 0.012 & 0.048 \\
Lower Mantle & 0.012 & 0.048 \\
\hline
\end{tabular}
\caption{\label{tab1} U and Th mass abundances in different layers of  
the Earth's crust and mantle used in this work~\cite{Enomoto-thesis-2005}.  
In the Earth's core these mass abundances are assumed to be zero.}
\end{table} 
\begin{figure}[ht]
\begin{center}
\vglue -0.5cm
\hglue -0.3cm
\includegraphics[width=0.50\textwidth]{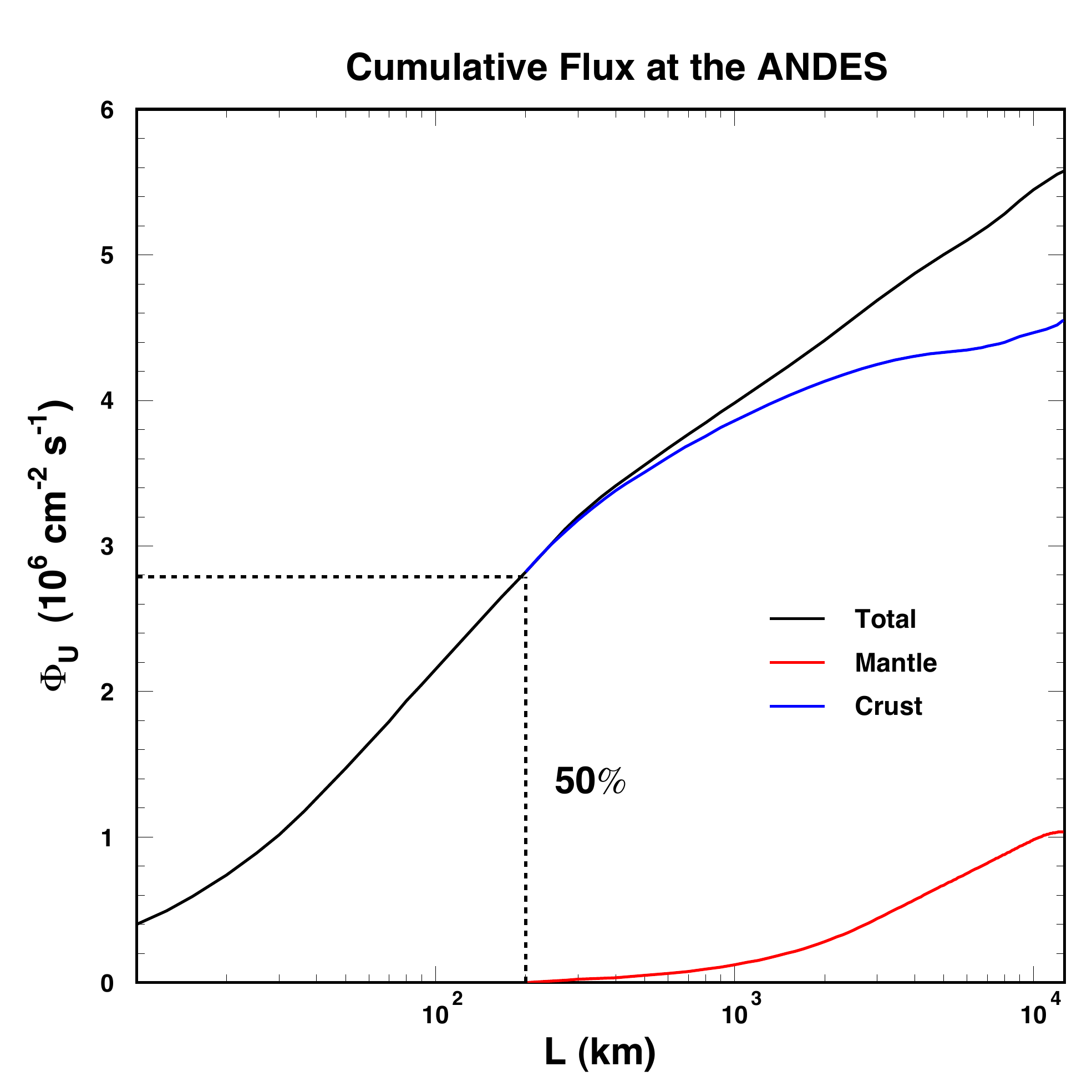}
\end{center}
\vglue -0.7cm
\caption{Cumulative flux of geoneutrinos coming from 
the decay of $^{238}$U as a function of the distance 
from the ANDES neutrino detector. }
\label{fig:geonu-flux-distance}
\end{figure}

We have computed the expected total fluxes of $\bar{\nu}_e$ at ANDES 
coming,  respectively, from U and Th to be, 
$\Phi_{\rm U} = 5.58 \times 10^{6}$ cm$^{-2}$ s$^{-1}$ ($3.04 \times
10^{6}$ cm$^{-2}$ s$^{-1}$) and $\Phi_{\rm Th} = 4.78 \times 10^{6}$
cm$^{-2}$ s$^{-1}$ ($2.60 \times 10^{6}$ cm$^{-2}$ s$^{-1}$) without
neutrino oscillation (with oscillation).  In
Fig.~\ref{fig:geonu-flux-distance} we present the geoneutrino
cumulative flux for the U chain as a function of the distance from the
ANDES laboratory. We observe that 50\% of the flux comes from $\simeq$ 
200 km from the detector and about 20\% of the flux comes from the mantle. 

\begin{figure}[!t]
\begin{center}
\vglue -1.0cm
\hglue -0.3cm
\includegraphics[width=0.50\textwidth]{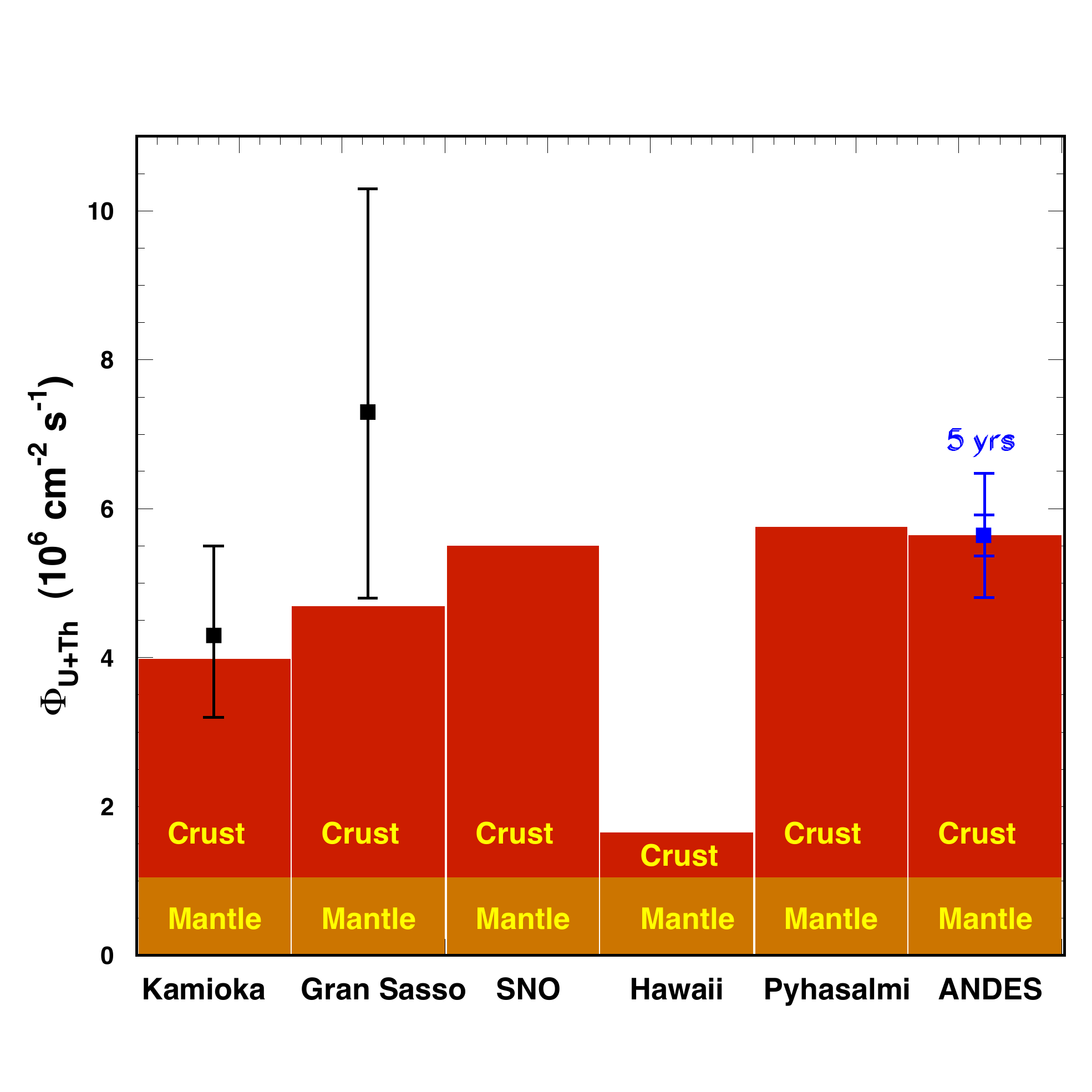}
\end{center}
\vglue -1.0cm
\caption{Total geoneutrino flux (oscillated) expected at Kamioka, 
Gran Sasso, SNO, Hawaii, Pyh\"asalmi and ANDES. We show the expected
  contribution from crust and mantle, in each case, as well as
  KamLAND~\cite{KamLAND-nature-geo2011} and Borexino
  data~\cite{Bellini:2010hy} points. We also show (in blue) the precision 
  of the expected measurement by ANDES after 5 years at 1 and 3 $\sigma$ CL.}
\label{fig:geonu-flux-exp}
\end{figure}

In Fig.~\ref{fig:geonu-flux-exp} we show our expectations for the
total oscillated geoneutrino flux at Kamioka, Gran Sasso, SNO, Hawaii,
Pyh\"asalmi and ANDES, discriminating the crust and mantle
contributions in each case. Pyh\"asalmi in Finland is a possible site
for the proposed 50 kt LENA neutrino detector~\cite{Wurm:2011zn}.  We
also show KamLAND~\cite{KamLAND-nature-geo2011} and Borexino
data~\cite{Bellini:2010hy} points to compare with the precision of the
expected measurement by ANDES after 5 years of data taking.  According
to Ref.~\cite{Fiorentini:2012yk} the current KamLAND and Borexino
results combined imply the geoneutrinos from the mantle have been
observed at 2.4 $\sigma$ CL. Clearly ANDES by itself, after 5 years,
is able to establish the mantle geoneutrino component at a level of
about 3 $\sigma$ or better.

Let us now discuss the expected number of geoneutrino induced events
at the ANDES neutrino detector.  For one year of operation ($3 \times
10^{7}$ s) and 80\% detector efficiency, we have calculated the total
number of geonetrinos expected at the ANDES reference detector to be
82.4 (64.8 from U, 17.6 from Th).  About 16 of these events would be
from the mantle and 35 events would have $E_\nu >$ 2.3 MeV, coming
exclusively from the U chain.  To illustrate the site dependence we
show in Table~\ref{tab2} our estimation for the corresponding number
of geoneutrino events in different locations assuming the same
reference detector (the same number of free protons, efficiency and
exposure).  From this table we see that the expected number of
geoneutrino events at the ANDES location is comparable to SNO 
and to Pyh\"asalmi.

\begin{table}
\begin{tabular}{|l|c|c|c|}
\hline
Location & Number from U & Number from Th & Total\\
\hline
Gran Sasso & 53.8 & 14.7 & 68.5  \\
Kamioka    &  45.7 & 12.4 & 58.1 \\
Hawaii &  18.5 & 5.0 & 23.5 \\
Sudbury & 63.2 & 17.2 & 80.4 \\
Pyh\"asalmi &  66.1 & 18.0 & 84.1 \\
ANDES &64.8  & 17.6  & 82.4 \\
\hline
\end{tabular}
\caption{\label{tab2} 
Expected number of geoneutrino events for our reference 
3 kt liquid scintillator detector operating during a year with 80\% efficiency 
 at different locations.}
\end{table} 

Such a detector operating during 10 years could accumulate more than 
800 geonetrino events (160 from the mantle alone), allowing  not only for 
a better determination of U and Th mass abundances in the crust and mantle  
but also for the investigation of their presence in  the Earth's core.
Clearly if an even larger detector, say 10 kt, could be envisaged the  
scientific reach could be even more significant.

\section{Supernova Neutrinos}
\label{sec:SN-neutrinos}

\subsection{Preliminaries}
\label{subsec:preliminaries}

If we consider the entire universe, the SN event rate is not so low,
several SN explosions per second. However, if we restrict to our
galaxy, more interesting in terms of SN neutrino observations, the
estimated event rate of such nearby SN drops down to about $\lsim 3$ per
century~\cite{Bergh:1991ek,Tammann:1994ev,Diehl:2006cf}.
In fact, in last $\sim$ 30 years, since when the Baksan neutrino
detector started to operate in 1980, only neutrinos from the 
explosion of SN1987A in the Large Magellanic Cloud, one of the satellite
galaxies of the Milky Way, were observed by the
Kamiokande~\cite{Hirata:1987hu}, IMB~\cite{Bionta:1987qt}, and
Baksan~\cite{Alekseev:1988gp} neutrino detectors.

Fortunately today, compared to that epoch, we are better prepared
for SN neutrino observations as there are much larger neutrino
detectors, such as Super-Kamiokande~\cite{Ikeda:2007sa} and
IceCube~\cite{Abbasi:2011ss}, which are currently in operation.
However, since the nearby SN rate is quite low, it is better to have
as many neutrino detectors as possible, to be ready for the next SN
event.  This will maximize the chance to obtain as much information as
possible on SN neutrinos, leading to a better understanding of the SN
explosion dynamics.

Since the last galactic SN, SN1604, was observed more than 400 years 
ago  and we do not know when exactly the next one will occur, it
is important to have neutrino detectors with larger running time. 
Having many neutrino detectors is also helpful in forming a
network like SNEWS~\cite{Antonioli:2004zb} that will readily alert the 
astronomers about the occurrence of a nearby SN event enabling them 
not to miss the initial phase of the time evolution of the SN light
curve.

It is theoretically expected~\cite{Bethe:1990mw} that almost all 
($\sim 99$ \%)  the energy released
by gravitational collapse is carried away in the form of neutrinos, 
which was consistent with the observed data of SN1987A neutrinos.
Roughly speaking, the neutrino emission from a SN explosion can be
divided into four periods: (i) the infall phase, which starts several
tens of ms before the bounce; (ii) the shock breakout neutronization
burst, which lasts up to a few tens of ms after the bounce; (iii)
the accretion phase, during from a few tens of ms up to several
hundreds of ms after the bounce and (iv) the Kelving-Helmholtz cooling
phase, during up to $\sim$ 10-20 s after the bounce.

Emission of $\nu_e$ starts during the infall phase, though the
luminosity is not yet so large.  In the neutronization burst phase,
there is a strong $\nu_e$ burst in a very short period of time, $\sim
10$ ms, and the emission of the other flavors, as well as of
$\bar{\nu}_e$, is suppressed.  During the accretion phase, the fluxes
of $\nu_e$ and $\bar{\nu}_e$ are expected to be significantly larger
than that of the other flavors and the energy hierarchy $\langle
E_{\nu_e} \rangle < \langle E_{\bar{\nu}_e} \rangle < \langle
E_{\nu_x} \rangle$ is expected.  Here we use the notation $\nu_x$ to
refer to any non-electron neutrino since for our purpose
they can be treated, in good approximation,  as a single species.
During the cooling phase the emission of all flavors of neutrinos and 
antineutrinos, with  similar luminosities, is predicted. 

It is considered that the energy spectra of SN neutrinos can be
 approximately described by Fermi-Dirac distributions with some
non-zero chemical potential, which is in general necessary to account
for the non-thermal feature of the SN neutrino spectra.
In  this work, for  any flavor,  $\nu_\alpha$, we
will  use  the  following  parameterization,  which is  based  on  the
numerical         simulations          by         the         Garching
group~\cite{Keil:2002in,Keil:2003sw,Buras:2002wt}, for the SN neutrino
spectra at the Earth in the absence of neutrino oscillation,

\begin{widetext}
\begin{equation}
  F^0_{\nu_\alpha}(E) = \frac{1}{4 \pi D^2}\ 
  \frac{\Phi_{\nu_\alpha}}{\vev{E_{\nu_\alpha}}}\,
  \frac{\beta_{\alpha}^{\beta_{\alpha}}}{\Gamma(\beta_{\alpha})}  
  \left[\frac{E}{\vev{E_{\nu_\alpha}}}\right]^{\beta_{\alpha}-1} 
 \exp\left[-\beta_{\alpha}\frac{E}{\vev{E_{\nu_\alpha}}}\right], \,
\label{eq:flux-Garching}
\end{equation}
\end{widetext}
where $D$ is the distance to the SN, $\Phi_{\nu_\alpha}$ is the total
number of $\nu_\alpha$ emitted, $\vev{E_{\nu_\alpha}}$ is the average
energy of $\nu_\alpha$ and $\beta_{\alpha}$ is a parameter which
describes the deviation from a thermal spectrum (pinching effect) that can
be taken to be $\sim 2-4$, $\Gamma(\beta_\alpha)$ is the gamma function. 

This parameterization seems to describe better 
the SN neutrino spectra obtained by numerical simulations.
We note, however, that our results would not change 
much even if we had used instead Fermi-Dirac distributions 
with a non-zero chemical potential.  
During the neutrino emission, as many SN simulations indicate, the
shape of $F^0_{\nu_\alpha}(E)$ is expected to change in time, which
means that the average neutrino energies as well as their luminosities are,
in general, functions of time. We do not explore this feature in
this work~\footnote[1]{Since the observation of SN events in the ANDES
  laboratory is likely to be statistically limited, as we will see in
  the next subsections, and also due to the SN model uncertainty on 
  the time dependence of $F^0_{\nu_\alpha}(E)$, a detailed time
  dependent analysis will not be considered here.}.

For the sake of discussion, unless otherwise explicitly stated, we use
the SN parameters summarized in Table \ref{table:SN-parameters} as 
our reference values.
We assume that the total energy released by SN neutrinos is 3$\times
10^{53}$ erg, which is equally divided by 6 species of active
neutrinos, and consider 10 kpc as a typical distance to the SN.
The chance that the actual distance could be even smaller, say 5 kpc, 
seems, however, to be not so small.  This can be seen in
Fig.~\ref{prob-SN-distance} in the Appendix \ref{appendix-shadowing},
where  the expected SN distributions as a function of the
distance from the Earth, based on the distribution models considered
in~\cite{Mirizzi:2006xx}, are shown.

\begin{table}[!h]
\begin{centering}
\begin{tabular}{c}
\hline \hline Reference SN parameters \\ \hline Distance to the SN : D
= 10 kpc \\ Spectra parmetrization in Eq.(\ref{eq:flux-Garching}) with
$\beta_\alpha$ = 4 for all flavors \\ $\langle E_{\nu_e} \rangle$ = 12
MeV ~~ $\langle E_{\bar{\nu}_e} \rangle$ = 15 MeV ~~ $\langle E_{\nu_x}
\rangle$ = 18 MeV \\
$E^{\text{tot}}_{\nu_\alpha} = 
\langle E_{\nu_\alpha} \rangle \Phi_{\nu_\alpha} = 
5 \times 10^{52}$ erg ~ for all flavor 
\tabularnewline
\hline \hline  
\end{tabular}
\par\end{centering}
\caption{Reference SN parameters to be used throughout this paper,
  unless otherwise explicitly stated.}
\label{table:SN-parameters}
\end{table}


Due to  neutrino oscillations in the SN envelope, 
the SN neutrino flux spectra at Earth are in
general expected not to be given by Eq.~(\ref{eq:flux-Garching}) but by 
flavor mixtures (superpositions) of them.
Without loss of generality, 
the flux spectrum of observable $\bar{\nu}_e$ at 
the Earth can be expressed as 
\begin{equation}
F_{\bar{\nu}_e} (E)
= \bar{p} (E)\,F_{\bar{\nu}_e}^0(E)
+[1-\bar{p} (E)] F_{\bar{\nu}_x}^0(E) \, ,
\label{eq:SN-flux-vac}
\end{equation}
where $F_{\bar{\nu}_e}^0(E)$ and 
$F_{\bar{\nu}_x}^0(E)$ are, respectively, 
the original spectra of $\bar{\nu}_e$ 
and $\bar{\nu}_{\mu,\tau}$ at the SN neutrinosphere 
in the absence of any oscillation effect. 
Eq.~(\ref{eq:SN-flux-vac}) implies that as long as the final
observable spectrum at Earth, $F_{\bar{\nu}_e} (E)$, is concerned,
thanks to the very similar spectra of $\bar{\nu}_\mu$ and
$\bar{\nu}_\tau$ in the SN, we can work within the effective 2 flavor
mixing scheme.
Note that if there is some difference between $\bar{\nu}_\mu$ and
$\bar{\nu}_\tau$ at the origin, $F_{\bar{\nu}_x}^0(E)$ must be given
by $c_\mu F_{\bar{\nu}_\mu}^0(E) + c_\tau F_{\bar{\nu}_\tau}^0(E)$
where the values of $c_\mu$ and $c_\tau$, which satisfy $c_\mu +
c_\tau = 1$, depend on the mixing and oscillation scenario.

The function $\bar{p}(E)$ is the $\bar{\nu}_e$ survival probability. 
This is, in general, a function of the neutrino energy $E$ and incorporates
all oscillation effects inside the SN, including the standard
Mikheyev-Smirnov-Wolfenstein (MSW) effect~\cite{Dighe:1999bi}, the 
so called collective effects~\cite{collective-effects} as well as other 
possible effects coming from shock wave~\cite{SN-shockwave}, density
fluctuations~\cite{Fogli:2006xy}, turbulence~\cite{Friedland:2006ta},
etc. Note that $\bar{p}(E)$ can also depend on time. 

From this expression, one can see that for a given value of
$\bar{p}(E) \ne 1$, a larger difference between $F_{\bar{\nu}_e}^0(E)$
and $F_{\bar{\nu}_x}^0(E)$ corresponds to a larger observable effect
of SN neutrinos at the Earth.  Clearly, there is no observable effect if
$F_{\bar{\nu}_e}^0(E) =F_{\bar{\nu}_x}^0(E)$.

According to recent SN
simulations~\cite{Fischer:2009af,Huedepohl:2009wh,Fischer:2011cy},
especially during the cooling phase, the mean energies of the
different flavors and their luminosities are likely to be similar.
If so, as mentioned above, any observable oscillation effect tends to
vanish.  Nevertheless, we should keep in mind that the SN neutrino
spectra must be ultimately determined by the future SN neutrino
observations, independently from any theoretical predictions of 
numerical SN simulations.

If only the standard MSW effect in the SN envelope is operative, then
it is straightforward to calculate the value of $\bar{p}(E)$ for a
given value of $\theta_{13}$ and a fixed mass
hierarchy~\cite{Dighe:1999bi}.  Since we know now the value of
$\theta_{13}$ rather well thanks to the recent measurements by
accelerator~\cite{Abe:2011sj,Adamson:2011qu} and reactor
experiments~\cite{Abe:2011fz,An:2012eh,collaboration:2012nd} (see also
the combined analysis in Ref.~\cite{Machado:2011ar}), the only open
question in this scenario is the neutrino mass hierarchy.
This will be referred to as the {\it standard scenario}.

The presence of other effects beyond the standard scenario, such as 
collective oscillations, shock waves, turbulence, etc., 
can cause a significant modification to $\bar{p}(E)$,  in the 
case of the inverted mass hierarchy. 
However, some recent studies~\cite{Chakraborty:2011nf,Sarikas:2011am} 
indicate that as long as the accretion phase is concerned, 
collective effects should be strongly suppressed due to the 
high matter density. This implies that in this phase 
only the standard scenario, the usual MSW effect, would be 
operative. 

For the normal mass hierarchy, in the standard scenario,
neglecting corrections from other effects, we expect 
that $\bar{p}(E) \approx c_{12}^2$~\cite{Dighe:1999bi}, and therefore, 
\begin{equation}
F_{\bar{\nu}_e} (E)
\approx c^2_{12}F_{\bar{\nu}_e}^0(E)
+s^2_{12}F_{\bar{\nu}_x}^0(E).
\label{eq:SN-spec}
\end{equation}
On the other hand, for the inverted mass hierarchy, by taking into
account the recent observation of non-zero $\theta_{13}$ by
accelerator~\cite{Abe:2011sj,Adamson:2011qu} and
reactor~\cite{Abe:2011fz,An:2012eh} experiments, it is expected that
$F_{\bar{\nu}_e} (E) \approx
F_{\bar{\nu}_x}^0(E)$~\cite{Dighe:1999bi}, {\it i.e.,} $\bar{p}(E) \to
0$, due to the adiabatic conversion inside the SN driven by the mass
squared difference $\Delta m^2_{32}$, relevant to atmospheric
neutrinos.
Hence, it is expected that independently of the mass hierarchy, even
after taking into account possible corrections due to collective
oscillations, shock waves, density fluctuations, etc., the value of
$\bar{p}(E)$ in Eq.~(\ref{eq:SN-flux-vac}) satisfy $0 \le \bar{p}(E)
\le c^2_{12} \simeq 0.69$~\cite{Lunardini:2009ya}.

\begin{figure}[!t]
\begin{center}
\hglue -0.4cm
\includegraphics[width=0.70\textwidth]{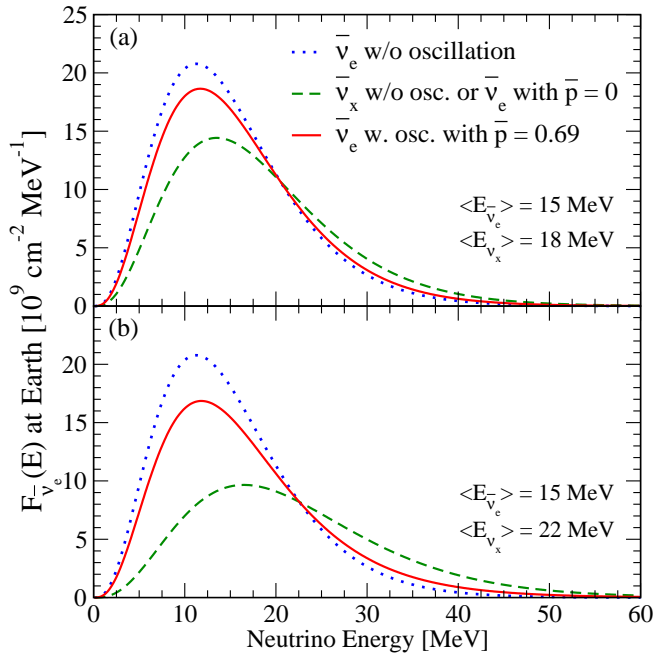}
\end{center}                          
\vglue -0.8cm
\caption{The expected neutrino flux spectra at the Earth for a SN
  located at 10 kpc from the Earth.  To calculate SN neutrino flux at
  the Earth, we assumed that at the SN neutrinosphere, $\langle
  E_{\bar{\nu}_e} \rangle $ = 15 MeV (dotted blue curves) whereas
  $\langle E_{\nu_x} \rangle $ = 18 MeV and 22 MeV for the upper and
  lower panel, respectively (dashed green curves).  For both
  cases, we show the expected $\bar{\nu}_e$ flux at the Earth by the
  solid red curves, given by Eq.~(\ref{eq:SN-spec}).  For
  simplicity, we consider that only the standard MSW effect plays  
  a role in oscillations so that $\bar p=$ 0.69 (0) corresponds to 
  the normal (inverted) mass hierarchy.}
\label{fig:SN-spec-10kpc}
\end{figure}

In Fig.~\ref{fig:SN-spec-10kpc} we show the expected neutrino flux
spectra at the Earth for a SN located at 10 kpc from the Earth, using
the typical average SN neutrino energies we consider in this work,
$\langle E_{\bar{\nu}_e} \rangle $ = 15 MeV and $\langle E_{\nu_x}
\rangle $ = 18 MeV and 22 MeV for the upper and lower panels,
respectively.  The expected $\bar{\nu}_e$ flux spectra at the Earth
are shown by the solid red curves for the case of normal mass
hierarchy which is given by Eq.~(\ref{eq:SN-spec}), whereas for the
inverted mass hierarchy, $\bar{\nu}_e$ spectra is given by the dashed
green curves.
We note that since we assumed that the total SN neutrino luminosity
was equally divided into the 6 species of neutrinos, a larger average energy
implies a smaller flux, as we can see from the dashed green curves in
Fig.~\ref{fig:SN-spec-10kpc}.

\begin{figure}[!t]
\begin{center}
\hglue -0.4cm
\includegraphics[width=0.70\textwidth]{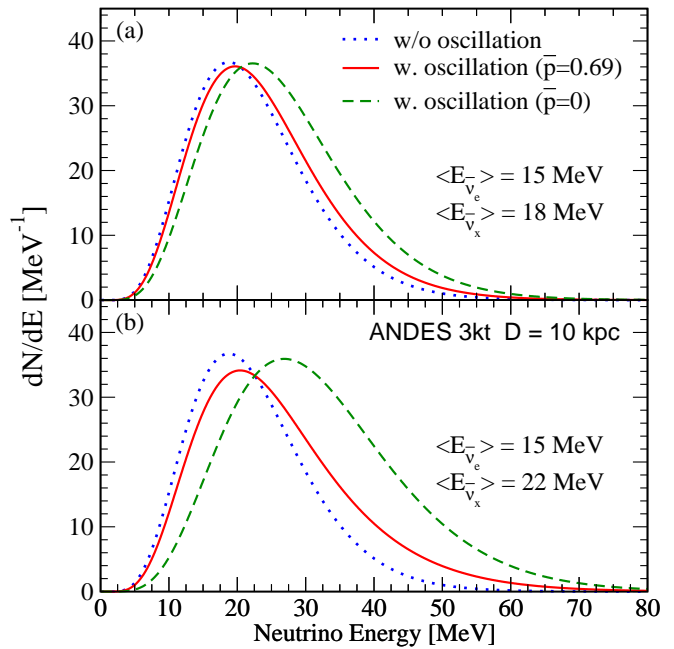}
\end{center}                          
\vglue -0.8cm
\caption{Expected energy distribution of $\bar \nu_e$ 
events coming from a SN located at 10 kpc from the Earth at 
the ANDES 3 kt detector. 
We assumed that at the SN neutrinosphere, 
$\langle E_{\bar{\nu}_e} \rangle $ = 15 MeV (dotted blue curves) 
whereas $\langle E_{\nu_x} \rangle $ = 18 MeV and 22 MeV 
for the upper and lower panels, respectively (dashed green curves). 
For both cases, we show the expected $\bar{\nu}_e$ flux at 
the Earth  for the normal (inverted) mass hierarchy 
by the solid (dashed) red curves.}
\label{fig:SN-event-10kpc}
\end{figure}

\subsection{Inverse beta decay reaction}
\label{subsec:inverse-beta}

In this paper, we focus on two main reactions. One is the
inverse beta decay $\bar{\nu}_e + p \to n + e^+$, which is the predominant  
channel due to its larger cross section, and the other is the
proton-neutrino elastic scattering, $\nu + p \to \nu + p$, which is
useful to determine the original  $\nu_x$
spectra~\cite{Beacom:2002hs,Dasgupta:2011wg}.
Based on the fluxes shown in Fig.~\ref{fig:SN-spec-10kpc}, we present
in Table \ref{table:SN-events} the expected number of events we have
computed for the ANDES neutrino detector in the absence and presence
of neutrino oscillations, for normal and inverted mass hierarchy and
different $\langle E_{\nu_x} \rangle$.

Only for Table \ref{table:SN-events}, for the purpose 
of comparison, we have considered three different chemical compositions for  
the liquid scintilator. These are mixtures that are already used by the 
existing or planned detectors, KamLAND~\cite{KamLAND}, 
Borexino~\cite{Alimonti:2000xc} and SNO+~\cite{Kraus:2010zzb}. 
KamLAND is based on the mixture of 80\% of C$_{12}$H$_{26}$ and 20\% 
of C$_{9}$H$_{12}$, whereas Borexino and SNO+ are based on
C$_{9}$H$_{12}$ (pseudocumene) and C$_{6}$H$_{5}$C$_{12}$H$_{25}$
(alkyl benzene), respectively.
For the rest of the paper we assume the same composition of SNO+. 

Regarding the oscillation probabilities, it is sufficient at this
point to consider the two extreme cases of the standard scenario where
$\bar{p} = 0.69$ and $\bar{p} = 0$ corresponding, respectively, to the
normal and inverted mass hierarchy. As mentioned before, apart from
some possible non-trivial energy dependence, one expects the actual
number of events to lay between these two cases even if other oscillation 
effects come into play.

\begin{table*}[!t]
\begin{centering}
\begin{tabular}{|c|c|c|c|c|}
\hline & \multicolumn{3}{c|} {Chemical Composition of the 
  Scintillator}&\tabularnewline \hline Reaction &\ (a)
C$_{12}$H$_{26}$ + C$_{9}$H$_{12}$\ &\  (b) C$_{9}$H$_{12}$ \ & \ (c)
C$_{6}$H$_{5}$C$_{12}$H$_{25}$ \ & Assumptions \\ &\ ( 80\% + 20\% ) &
\ \ pseudocumene\ \ \  & alkyl benzene & \tabularnewline \hline $\bar{\nu}_e + p
\to n + e^+$ & 873 & 630 & 762 & No Oscillation \tabularnewline \hline
\hline
$\bar{\nu}_e + p \to n + e^+$ & 924 & 669 & 804 & $\bar{p} = c_{12}^2$
= 0.69 (NH), $\langle E_{\nu_x} \rangle$= 18 MeV \tabularnewline
$\bar{\nu}_e + p \to n + e^+$ & 1038 & 750 & 903 & $\bar{p} $
= 0.0 (IH), \ \ $\langle E_{\nu_x} \rangle$= 18 MeV \tabularnewline
\hline 
$\bar{\nu}_e + p \to n + e^+$ & 957 & 690 & 834 & $\bar{p} = c_{12}^2$
= 0.69 (NH), $\langle E_{\nu_x} \rangle$= 20 MeV \tabularnewline
$\bar{\nu}_e + p \to n + e^+$ & 1140 & 825 & 993 & $\bar{p} $
= 0.0 (IH), \ \ $\langle E_{\nu_x} \rangle$= 20 MeV \tabularnewline
\hline 
$\bar{\nu}_e + p \to n + e^+$ & 987 & 714 & 858 & $\bar{p} =
c_{12}^2$ = 0.69 (NH), $\langle E_{\nu_x} \rangle$= 22 MeV
\tabularnewline $\bar{\nu}_e + p \to n + e^+$ & 1239 & 894 & 1080 &
$\bar{p}$ = 0.0 (IH), \ \ $\langle E_{\nu_x} \rangle$= 22
MeV \tabularnewline \hline \hline 
$\nu + p \to \nu + p$ & 294 & 318 & 453
&\ all flavors \ $T_{\text{que}} > 0.2$ MeV,\ $\langle E_{\nu_x} \rangle$= 18 MeV
\tabularnewline \hline
$\nu + p \to \nu + p$ & 399 & 405 & 561
&\ all flavors \ $T_{\text{que}} > 0.2$ MeV, \ $\langle E_{\nu_x} \rangle$= 20 MeV
\tabularnewline \hline
$\nu + p \to \nu + p$ & 510 & 492 & 663
&\ all flavors \ $T_{\text{que}} > 0.2$ MeV, \ $\langle E_{\nu_x} \rangle$= 22 MeV
\tabularnewline \hline
\end{tabular}
\par\end{centering}
\caption{Expected number of SN neutrino induced events for the inverse
beta decay and proton-neutrino elastic scattering for the 3 types of
liquid scintillators with the fiducial mass of 3 kt for a SN at 
10 kpc from the Earth. 
They are (a) 80\% of C$_{12}$H$_{26}$ and 20\% of C$_{9}$H$_{12}$ 
used for KamLAND, (b) C$_{9}$H$_{12}$ (pseudocumene) used 
for Borexino, and (c) C$_{6}$H$_{5}$C$_{12}$H$_{25}$ (alkyl benzene)
to be used for SNO+. 
NH and IH indicate the normal and inverted mass hierarchies,
respectively.  For $\nu + p \to \nu + p$, we considered the kinetic
(quenched) energy of the recoil proton, $T_{\text{que}}$, larger than
0.2 MeV following Refs.~\cite{Beacom:2002hs,Dasgupta:2011wg}.  }
\label{table:SN-events}
\end{table*}

In Fig.~\ref{fig:SN-event-10kpc} we show the expected $\bar{\nu}_e$
event number distribution $dN/dE$ for our reference ANDES neutrino
detector in the absence and presence of neutrino oscillations. 
Depending on the oscillation probabilities, 
we expect to have $\sim 800-1000$ events for 3 kt
(see Table \ref{table:SN-events}). 
In the presence of oscillation, the energy spectrum of 
$\bar{\nu}_e$ gets harder whereas its total flux will 
decrease, as we assume that the original value 
of $\Phi_{\nu_\alpha} \langle E_{\nu_\alpha} \rangle$ is constant for all
6 species. Nevertheless, the oscillation effect 
makes the expected observed event number larger, as 
the cross section depends on $\sim E_\nu^2$, which
overcomes the reduction of the flux due to oscillations. 

Due to neutrino oscillations the observed spectra at the 
Earth is, in general, a mixture of the original ones it would 
be desirable to be able to reconstruct the original spectra 
from data. 

In principle, by fitting the observed positron spectrum of 
$\bar{\nu}_e$ induced events at the detector, one can try to reconstruct
the original spectra (luminosites and average energies) 
of $\bar{\nu}_e$ and $\bar{\nu}_x$ as done, for example, 
in Refs.~\cite{Minakata:2001cd,Skadhauge:2006su}
but in practice, this may not be so trivial to do,   
even for a much larger detector such as Hyper-Kamiokande~\cite{Abe:2011ts}, 
due to the possible presence of degeneracies of 
the SN parameters~\cite{Minakata:2008nc}
or some unexpected large deviation of SN neutrino spectra 
from what is  usually assumed. 

As far as the reconstruction of the spectra of $\bar{\nu}_x$ 
is concerned, the use of the proton-neutrino elastic 
scattering seems to be more promising~\cite{Beacom:2002hs,Dasgupta:2011wg}.
This will be discussed in the next subsection. 
We propose to combine both, the inverse beta decay and 
proton-neutrino reactions, in order to identify the 
oscillation effect, or determine SN parameters 
if oscillation effect is known, 
as we will discuss in subsection~\ref{subsec:CC-vs-NC}.

\subsection{Proton-neutrino elastic scattering}
\label{subsec:proton-neutrino}

In this section, we focus on the proton-neutrino elastic 
scattering discussed in Refs.~\cite{Beacom:2002hs,Dasgupta:2011wg,Wurm:2011zn}. 
Since the proton-neutrino elastic scattering, 
$\nu + p \to \nu + p$, for all flavors, 
occurs via neutral current interactions, 
one can measure the total SN neutrino 
flux above a certain energy threshold, without worrying 
about any oscillation effect among active flavors, 
in a similar manner as the SNO experiment was able to measure the total solar 
neutrino flux of active flavors~\cite{Ahmad:2002jz}. 
If the average energies of $\nu_e$ and $\bar{\nu}_e$ are significantly
lower than that of $\nu_x$, then by counting the total number of
events above a certain recoil proton energy, one can measure the total
flux (luminosity) of $\nu_x$ flavor with reasonably 
good precision~\cite{Beacom:2002hs,Dasgupta:2011wg}.
If the energy spectra of SN $\nu_e$, $\bar{\nu}_e$, and $\nu_x$ are rather
similar, one can then try to determine the total neutrino flux.

To compute the number of events induced by the $\nu$-$p$ elastic
scattering, we follow the analysis procedure described in
Refs.~\cite{Beacom:2002hs,Dasgupta:2011wg}. 
We have checked that by using the same
information and assumption used in these references 
we could obtain results which
are in good agreement with the ones presented in these works. 

\begin{figure}[!h]
\begin{center}
\vglue -1.2cm
\hglue -0.3cm
\includegraphics[width=0.63\textwidth]{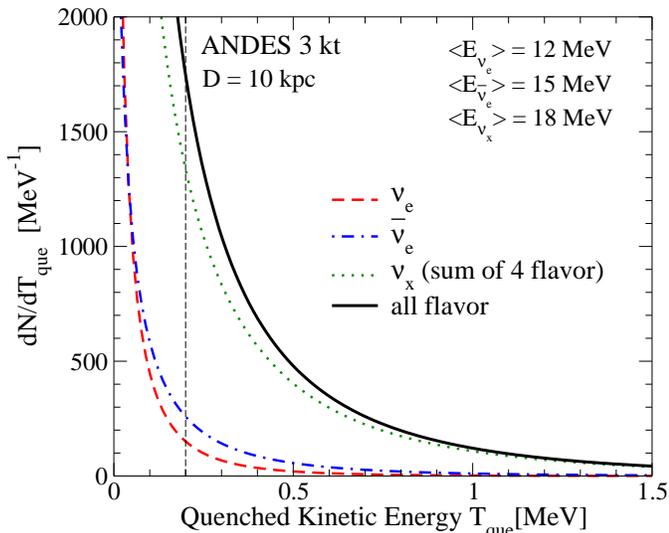}
\end{center}
\vglue -1.2cm
\caption{Distribution of events as a function of the quenched kinetic
  energy of the proton for the ANDES reference detector using the
  reference SN parameters defined in Table
  \ref{table:SN-parameters}. }
\label{fig:SN-nu-proton-ANDES}
\end{figure}

In Fig.~\ref{fig:SN-nu-proton-ANDES} we show the expected event number
distribution, $dN/dT_{\text{que}}$, as a function of the quenched
kinetic proton energy, $T_{\text{que}}$, for the ANDES reference
neutrino detector and for the reference SN parameters summarized in
Table \ref{table:SN-parameters}.
%

Following~\cite{Dasgupta:2011wg}, the energy range 
of $T_{\text{que}} > 0.2$ MeV was considered mainly 
because of the backgrounds coming from radioactive decays 
in the scintillator and surroundings, 
such as the one coming from the $\beta$ decay of $^{14}$C. 
The qualitative behavior of $dN/dT_{\text{que}}$ for the ANDES detector 
is very similar to the ones obtained in \cite{Dasgupta:2011wg}
for KamLAND, Borexino and SNO+ detectors (see Figs. 1-3 of
this reference), the main difference is the total number of events.

We have studied how accurately one can reconstruct the original neutrino 
flux by the energy distribution of recoil protons 
in the ANDES detector. 
As demonstrated explicitly in Ref.~\cite{Dasgupta:2011wg}
by considering the quenched proton recoil energy larger than 0.2 MeV, 
one can reconstruct the original neutrino flux 
above $\gsim $ 25 MeV.  The idea is to make an inversion of the calculation 
in order to produce the curves shown in Fig.~\ref{fig:SN-nu-proton-ANDES}. 

\begin{figure}[!h]
\begin{center}
\hglue -0.20cm
\includegraphics[width=0.52\textwidth]{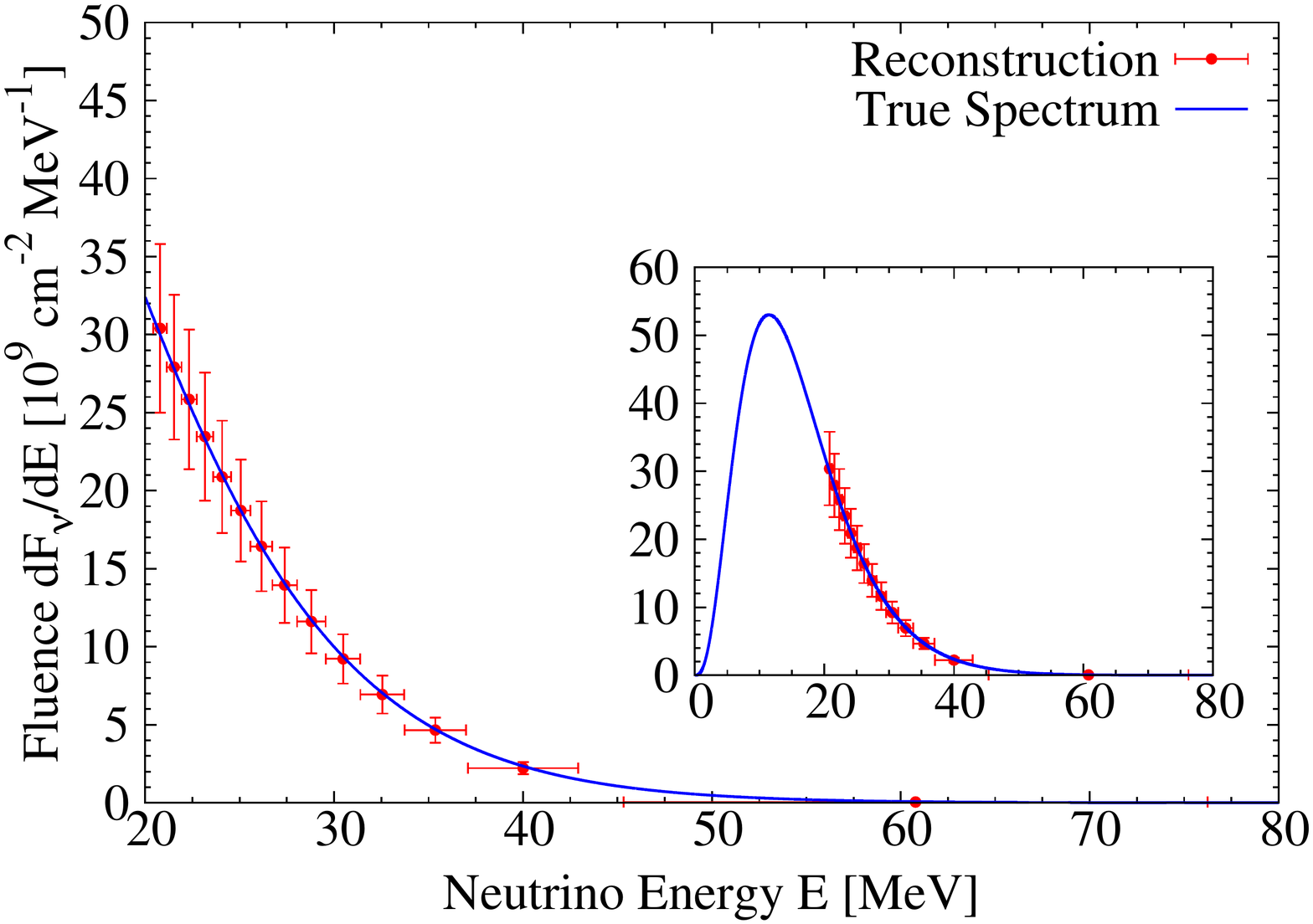}
\end{center}
\vglue -0.9cm
\caption{Reconstruction of the original SN neutrino flux (indicated 
by the red solid circles with error bar of 1 $\sigma$), 
from the simulated future data of proton-neutrino  
elastic scattering in the ANDES detector under 
the same assumptions used in Fig.~\ref{fig:SN-nu-proton-ANDES}, 
for the reference SN parameters defined in Table \ref{table:SN-parameters}.} 
\label{fig:SN-nu-reconstruction-ANDES}
\end{figure}

In Fig.~\ref{fig:SN-nu-reconstruction-ANDES} we show our result.  For
our reference SN, with the 3 kt ANDES detector, one can try to
determine the original neutrino flux with the precision of $\sim$ 15~\% for the neutrino energy $\sim$ 20-40 MeV.
Note that this result does not depend on the uncertainty 
on the effects of neutrino oscillations among 
active flavors, which could occur in the SN envelope. 
While the expected precision is not as good as the one
expected for the proposed much larger
LENA detector~\cite{Wurm:2011zn}, which has 50 kt fiducial volume,  
as we can compare our results with the one shown in 
Fig. 9 of Ref.~\cite{Dasgupta:2011wg}, 
the expected precision for the ANDES detector 
is better than the currently existing (planned) detectors like 
KamLAND and Borexino (SNO+). 

\begin{figure}[!h]
\begin{center}
\hglue -0.25cm
\includegraphics[width=0.53\textwidth]{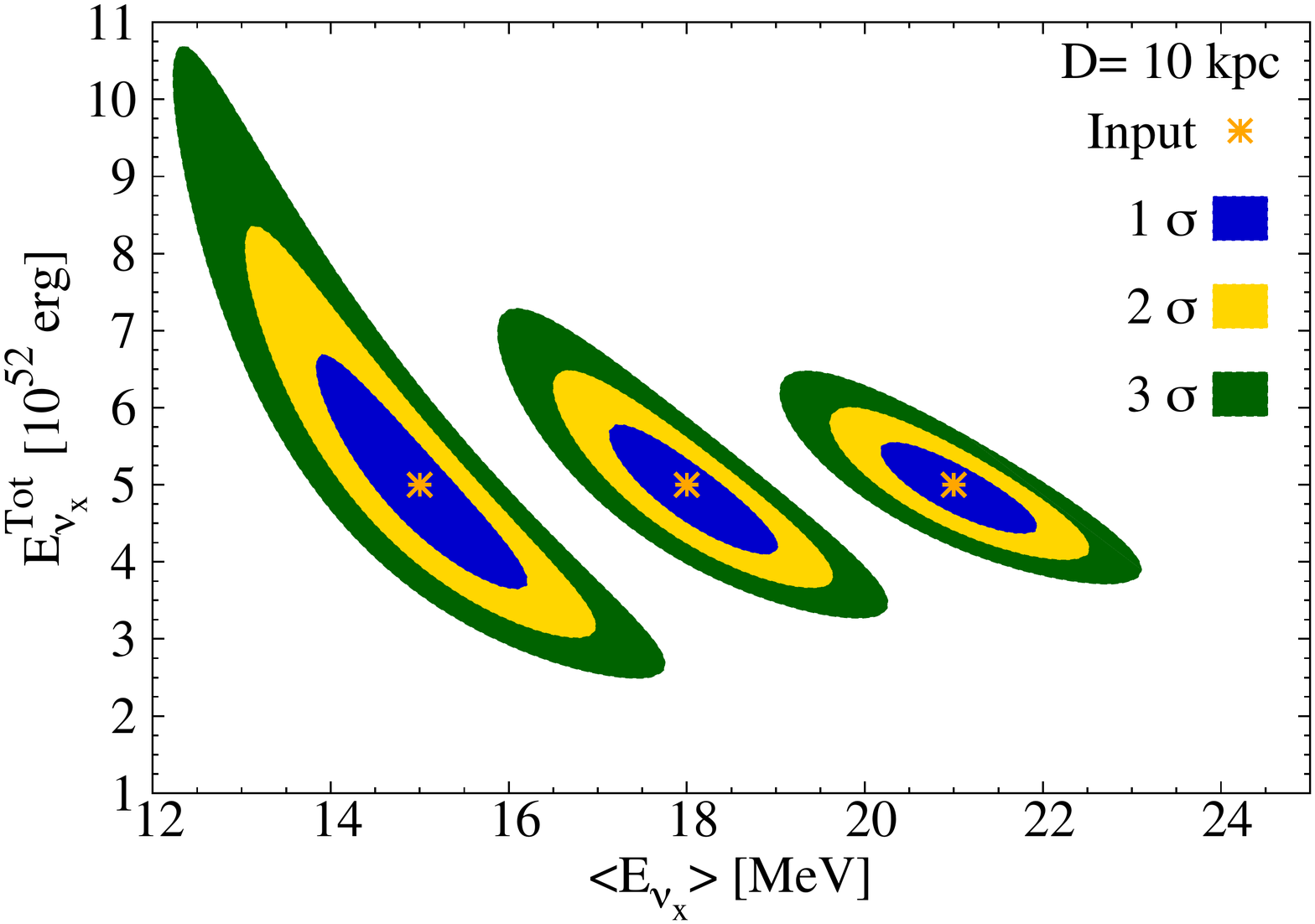}
\end{center}
\vglue -1.0cm
\caption{Sensitivity of the determination of 
the original $E^{\text{tot}}_{\nu_x}$ and $\langle E_{\nu_x} \rangle $ 
at 1,2 and 3 $\sigma$ CL, 
by proton-neutrino scattering events measured at the ANDES detector. 
Only the statistical uncertainties were taken into account. 
We consider three different input values for the average $\bar{\nu}_x$ 
energy,  $\langle E_{\nu_x} \rangle $ = 15 MeV,  18 MeV and 
21 MeV while that for $E^{\text{tot}}_{\nu_x}$ was fixed to be 
$5\times 10^{52}$ erg (they are indicated by the asterisk).
The other SN parameters are defined in Table \ref{table:SN-parameters}. 
}
\label{fig:proton-nu-Etot-E_nu_x}
\end{figure}

Next let us discuss with which precision we can try to determine the
original $\nu_x$ mean energy.  For this purpose, we perform a $\chi^2$
analysis by considering the input (true) values of SN parameters for
$\langle E_{\nu_x} \rangle $ = 15, 18 and 22  MeV and 
$E^{\text{tot}}_{\nu_x} = 5\times 10^{52}$ erg. 
We define $E^{\text{tot}}_{\nu_x}$ such that 
4$E^{\text{tot}}_{\nu_x}$ gives the total energy carried away 
by non-electron flavor neutrinos. 
For simplicity, we fix the other SN parameters for $\nu_e$ 
and $\bar{\nu}_e$ to our reference values, 
and only vary $E^{\text{tot}}_{\nu_x}$ and $\langle E_{\nu_x} \rangle $ 
in our fit, in order to have some feeling about the
precision we can achieve. 
We show our result in Fig.~\ref{fig:proton-nu-Etot-E_nu_x} where the
allowed parameter regions in the plane of $E^{\text{tot}}_{\nu_x}$ and
$\langle E_{\nu_x} \rangle $ is shown. Here only the statistical
uncertainties were taken into account.  From this, we conclude that
we can determine $\langle E_{\nu_x} \rangle $ and
$E^{\text{tot}}_{\nu_x}$ with the precisions of $\sim 6$\% (25\%) and
$\sim 20$\% (40\%), respectively, at 1 (3) $\sigma$ CL, 
for the case where the true values 
of $\langle E_{\nu_x} \rangle $ and 
$E^{\text{tot}}_{\nu_x}$ were assumed to be 18 MeV
and $5\times 10^{52}$ erg, respectively.

\subsection{Comparison of CC and NC induced events}
\label{subsec:CC-vs-NC}

As mentioned in the previous subsection, one of the important features
of the reaction induced by proton neutrino elastic scattering is that
it does not depend on the neutrino oscillation among active flavors,
as this reaction is induced by neutral current (NC).
On the other hand, the inverse beta decay is induced by charged
current (CC) interactions and so does depend on the neutrino
oscillation probability, as can be seen in Table
\ref{table:SN-events}.

Therefore, by comparing these two kinds of events, one can try to
infer, to some extent, the oscillation effect or the value of
$\bar{p}$ in Eq.(\ref{eq:SN-flux-vac}) in a less model dependent way.
Or conversely, if the mass hierarchy is known (from some other source, 
like one of the long-baseline neutrino oscillation experiments) 
by the time the next galactic SN occurs, and $\bar{p}$ can be predicted in
advance, one can try to determine better some other characteristic of
the SN neutrinos.  We observe that the comparison of CC and NC events was
also considered in \cite{Borexino-SN-nu}.

For the purpose of illustration, let us define 
the following ratio, 
\begin{equation}
R(N_{\bar{\nu}_e p}/N_{\nu p}) \equiv 
\displaystyle 
\left(\frac{ N_{\bar{\nu}_e p}}{N_{\nu p}}\right)^{\text{obs}},
\label{eq:Double-Ratio2}
\end{equation}
where $(N_{\bar{\nu}_e p}/N_{\nu p})^{\text{obs}}$ means
the ratio of the observed total number of events induced by 
the inverse beta decay and proton-neutrino elastic 
scattering.

\begin{figure}[!b]
\vglue -0.4cm
\begin{center}
\hglue -0.4cm
\includegraphics[width=0.53\textwidth]{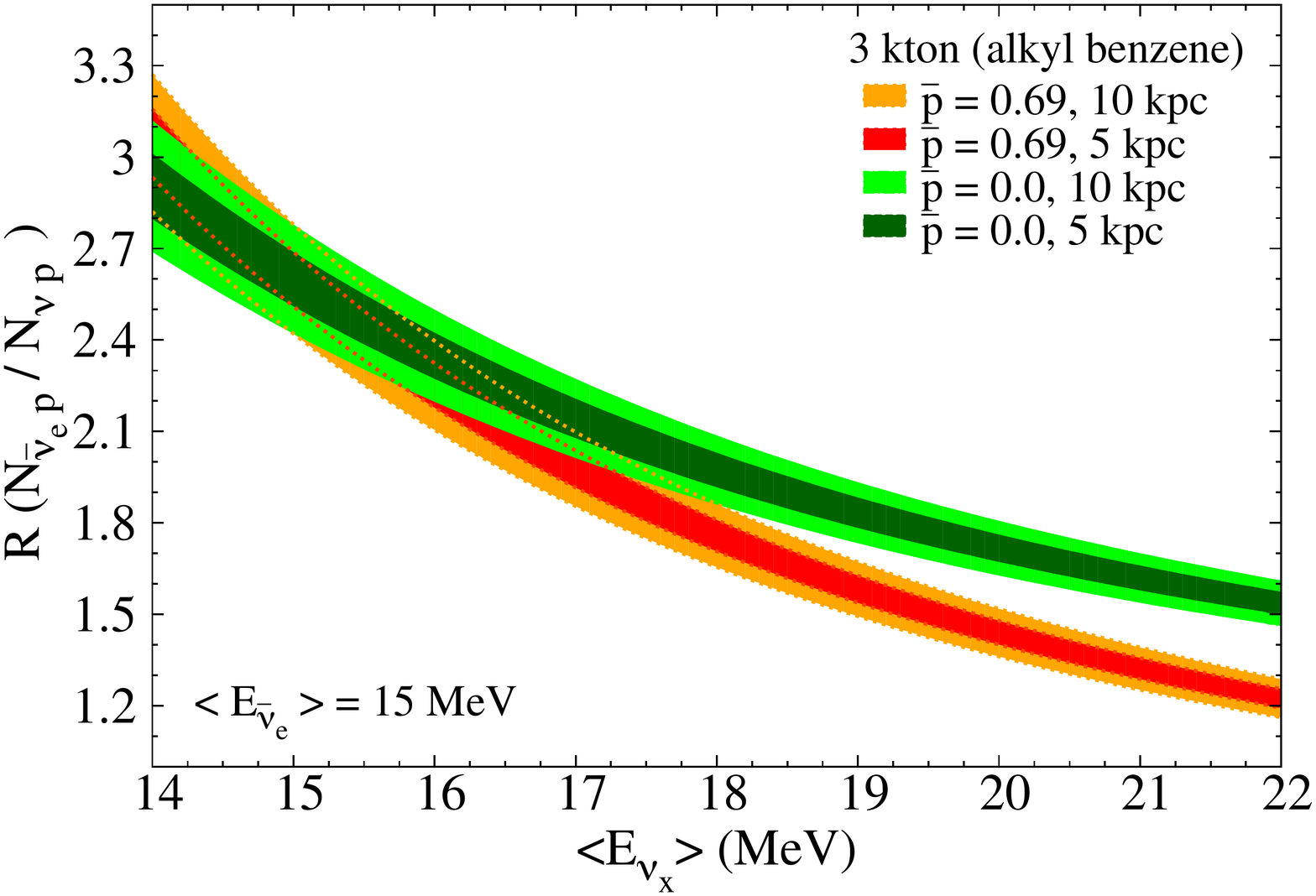}
\end{center}
\vglue -1.0cm
\caption{
The ratio, $R(N_{\bar{\nu}_e p}/N_{\nu p}) $, 
as a function of the true value of 
$\langle E_{\nu_x} \rangle$ for our reference  
ANDES neutrino detector.
The values of the SN parameters not specified in this plot are
the ones defined in Table \ref{table:SN-parameters}.
}
\label{fig:cc-nc-double-ratio-energy}
\end{figure}

In Fig.~\ref{fig:cc-nc-double-ratio-energy} we show this quantity as a
function of the true value of $\langle E_{\nu_x} \rangle$.  For
$\bar{p}$, we considered the two extreme values of the standard
  scenario, $\bar{p}$ = 0 and 0.69. 
For this study, we consider the two cases where 
$D =$  5 and 10 kpc, indicated by 
the darker and lighter color, respectively. 

We note that if $\langle E_{\nu_x} \rangle$ is significantly different
from $\langle E_{\bar{\nu}_e} \rangle$, one can try to identify the
presence of oscillation effect by inferring the value of $\bar{p}$
with some precision,    
which could be possible especially for the case
where the true value of $\langle E_{{\nu}_x} \rangle$ turns out to be
quite different from $\langle E_{\bar{\nu}_e} \rangle$.  
This may be the case during the accretion phase as energies and
luminosities of $\bar{\nu}_e$ and $\nu_x$ may be significantly
different, as recent SN simulations 
indicate~\cite{Fischer:2009af,Huedepohl:2009wh,Fischer:2011cy}.
Moreover, in this phase, as mentioned before recent
studies~\cite{Chakraborty:2011nf,Sarikas:2011am} 
indicate that the collective effects could be suppressed by
matter, which means that the neutrino conversion would be given by the
standard scenario. This would allows us to identify the oscillation
effect more easily, provided that the number of events for this phase
is large enough.

However, it must be stressed that differences or 
similarities between $\bar{\nu}_e$ and $\nu_x$ fluxes  
must be confirmed (or refuted) by observations. 
The real SN neutrino data 
could be very different from our expectations!
We also should keep in mind that 
the characteristics of SN neutrinos may depend
strongly on the SN (on the progenitor's  mass, chemical composition
of the core, etc.) (see e.g.,~\cite{Lunardini:2007vn}).

\begin{figure}[!b]
\vglue -0.4cm
\begin{center}
\hglue -0.3cm
\includegraphics[width=0.53\textwidth]{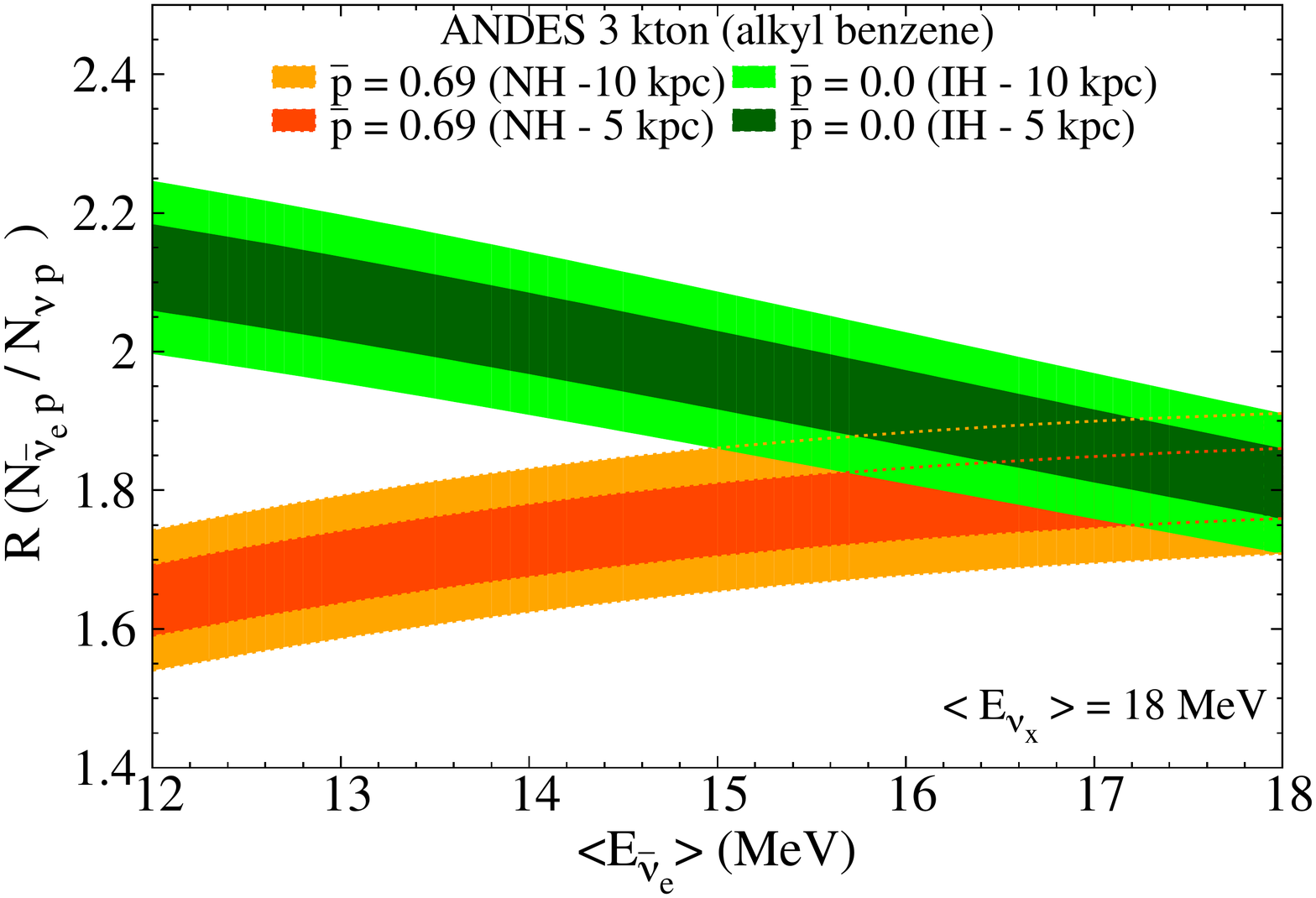}
\end{center}
\vglue -1.0cm
\caption{
Same as Fig.~\ref{fig:cc-nc-double-ratio-energy} but 
as a function of the true value of $\langle E_{\bar{\nu}_e} \rangle$. 
The values of the SN parameters not specified in this plot are
the ones defined in Table \ref{table:SN-parameters}.
}
\label{fig:cc-nc-double-ratio-energy2}
\end{figure}

In Fig.~\ref{fig:cc-nc-double-ratio-energy2} we show
the same quantity, $R(N_{\bar{\nu}_e p}/N_{\nu p})$, 
but as a function of the true (input) value of 
$\langle E_{\bar{\nu}_e} \rangle$
for the cases where the true value of 
$\langle E_{\nu_x} \rangle $ is 18 MeV. 
A similar conclusion can be drawn as before. Unless 
$\langle E_{\bar{\nu}_e} \rangle$ is significantly 
different from $\langle E_{\nu_x} \rangle $, 
it will not be very easy to distinguish 
$\bar{p} =0$ from $\bar{p} =0.69$. 
We can also use these results to do the opposite. Namely, if we know the
mass hierarchy by the time of the next galactic SN observation, we can
try to infer, to some extent, the original values of $\langle
E_{\bar{\nu}_e} \rangle$ and $\langle E_{\nu_x} \rangle $ from data.

\begin{figure}[!h]
\begin{center}
\vglue -0.4cm
\hglue -0.2cm
\includegraphics[width=0.50\textwidth]{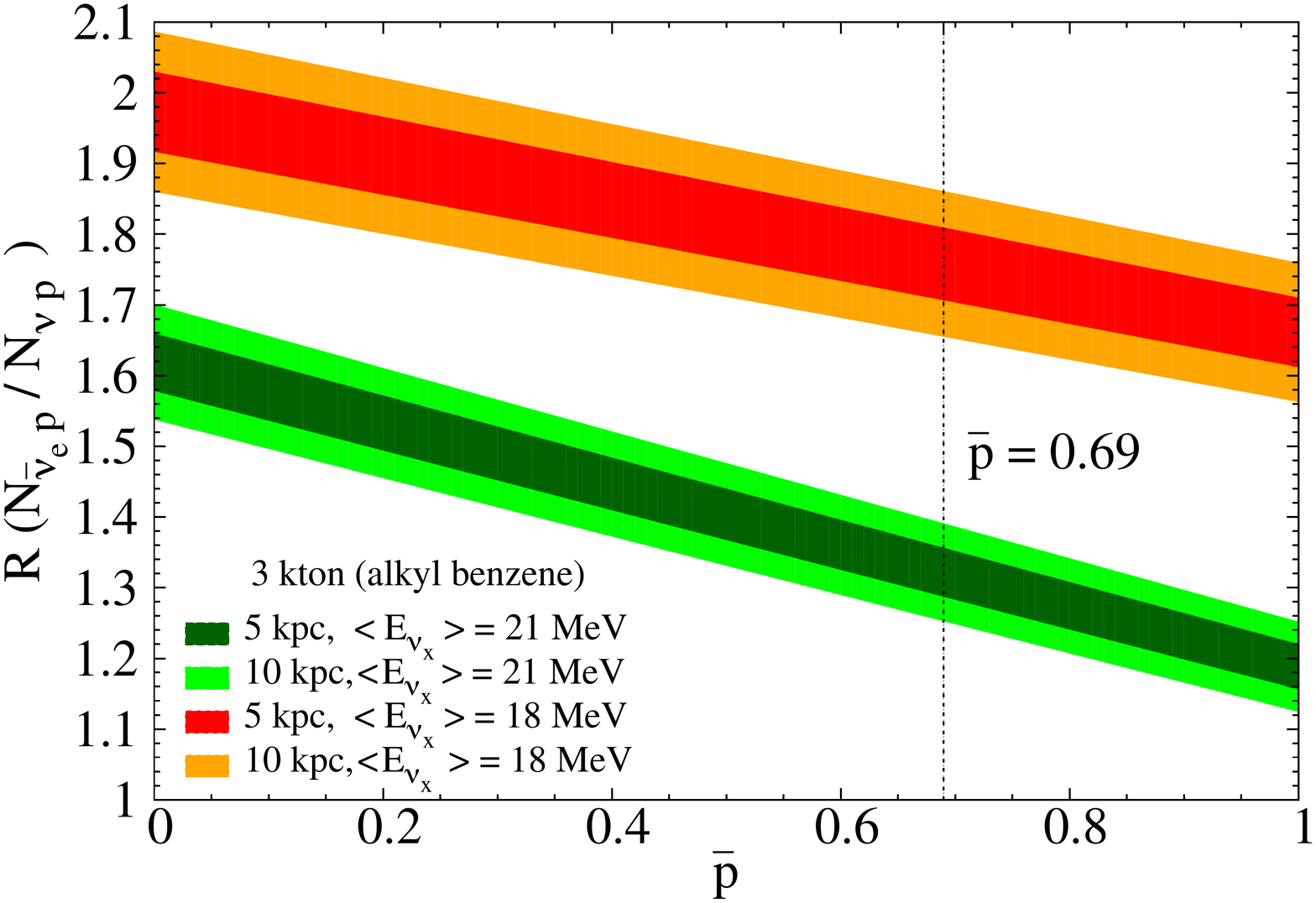}
\end{center}
\vglue -0.9cm
\caption{Same as Fig.~\ref{fig:cc-nc-double-ratio-energy} but as 
a function of $\bar{p}$ for the cases where the true value 
of $\langle E_{\nu_x} \rangle$ = 18 and 21 MeV. }
\label{fig:cc-nc-double-ratio-pbar}
\end{figure}

In Fig.~\ref{fig:cc-nc-double-ratio-pbar} we 
show $R(N_{\bar{\nu}_e p}/N_{\nu p})$ 
as a function of $\bar{p}$ 
for the cases where the true values of 
$\langle E_{\nu_x} \rangle $ are 18 and 21 MeV. 
We see again that unless $\langle E_{\nu_x} \rangle$ is
significantly different from $\langle E_{\bar{\nu}_e} \rangle$, it is
not easy to identify the oscillation effect, especially if the
hierarchy is normal and $\bar{p}$ = 0.69.  The case of inverted
hierarchy, $\bar{p}$ = 0.0, seems easier to be identified.

\begin{figure}[!h]
\begin{center}
\hglue -0.4cm
\includegraphics[width=0.50\textwidth]{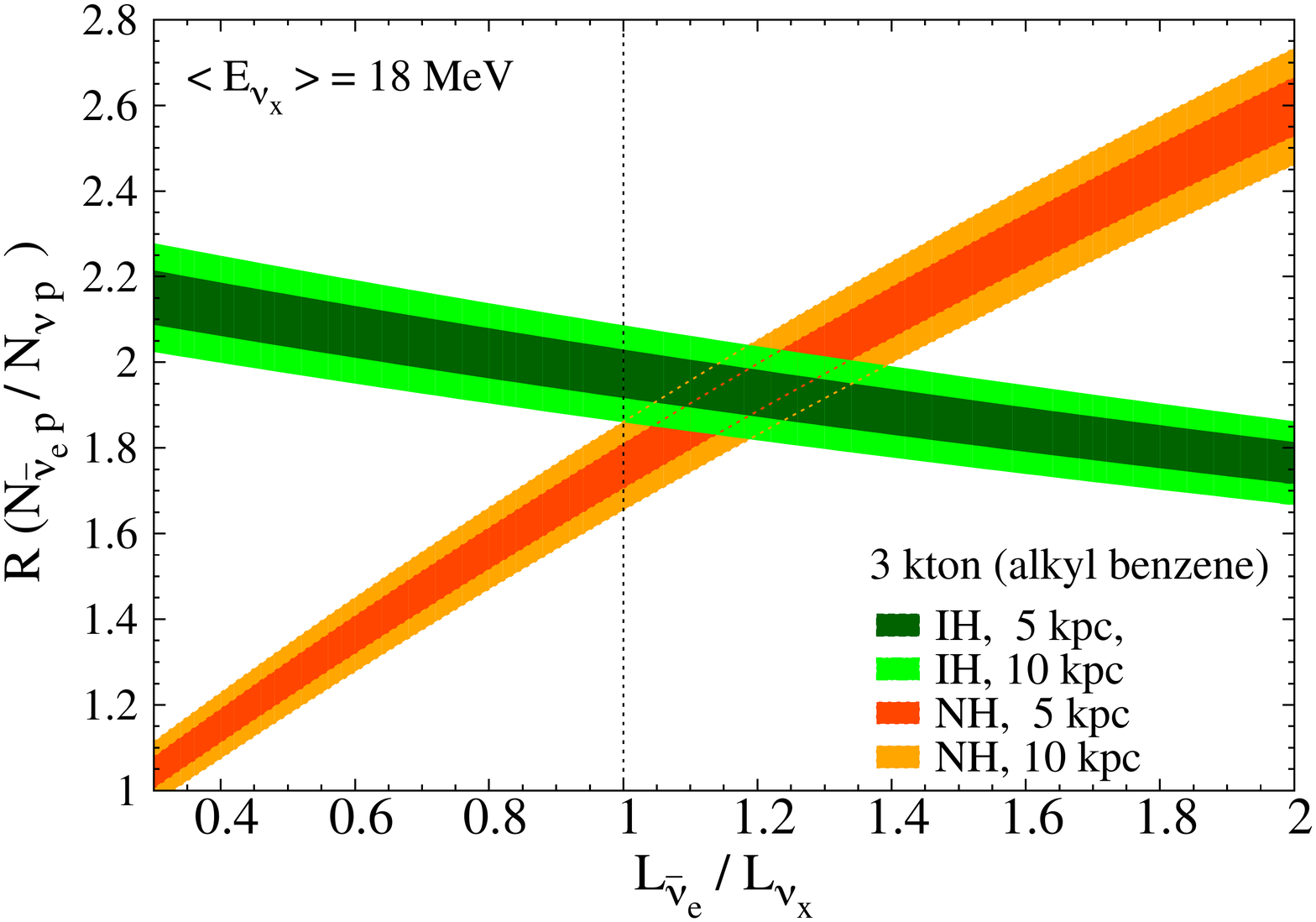}
\end{center}
\vglue -0.9cm
\caption{Same as Fig.~\ref{fig:cc-nc-double-ratio-energy} but as a
  function of the true value of $L_{\bar{\nu}}/L_{\nu_x} \equiv
  (\langle E_{\bar{\nu}_e} \rangle \Phi_{\bar{\nu}_e})/ (\langle
  E_{{\nu}_x} \rangle \Phi_{{\nu}_x})$ for the cases where the true
  value of $\langle E_{\nu_x} \rangle$ = 18 MeV.  The values of the
  SN parameters not specified in this plot are the ones defined in
  Table \ref{table:SN-parameters}.  }
\label{fig:cc-nc-double-ratio-luminosity}
\end{figure}

In Fig.~\ref{fig:cc-nc-double-ratio-luminosity} we show the ratio
$R(N_{\bar{\nu}_e p}/N_{\nu p})$ as a function of
$L_{\bar{\nu}}/L_{\nu_x} \equiv (\langle E_{\bar{\nu}_e} \rangle
\Phi_{\bar{\nu}_e})/ (\langle E_{{\nu}_x} \rangle \Phi_{{\nu}_x})$ for
the case where the true value of $\langle E_{\nu_x} \rangle$ is 18
MeV.  This plots indicate that if the luminosity of $\bar{\nu}_e$ is
larger than that of $\nu_x$ by $\sim 20$ \% or so, the cases of normal
and inverted mass hierarchy can be confused.  However, if
$\bar{\nu}_e$ luminosity is significantly larger or smaller (by $\sim 50$ \% or
more), then it is easier to distinguish the mass hierarchy.
On the other hand, if the mass hierarchy is known by the time the
next SN neutrinos are observed, then we can try to infer the difference
of the luminosities of $\bar{\nu}_e$ and $\nu_x$.

\subsection{Earth Matter Effect: Shadowing Probabilities}
\label{subsec:earth-matter}

One potentially interesting possibility is to observe the Earth
matter effect due to SN 
neutrinos~\cite{Dighe:1999bi,Lunardini:2001pb,Dighe:2003jg,Dighe:2003vm}.
If observed, it could unravel the neutrino mass hierarchy
and some properties of the SN neutrino fluxes.
Suppose that SN neutrinos are observed by more than two neutrino
detectors.  If some of them (but not all) receive SN neutrinos passing
through the Earth interior (shadowed by the Earth) but at the same
time, there are some other group of detectors which receive SN
neutrinos without passing through the Earth interior 
(non shadowed by the Earth), it would be interesting to compare the 
results of these two group of detectors.

Currently, several neutrino detectors, which can observe neutrinos
coming from a galactic SN, are in operation.  
Among them we can highlight Super-Kamiokande (SK)~\cite{Ikeda:2007sa}, 
KamLAND~\cite{KamLAND}, Borexino~\cite{Alimonti:2000xc},  
IceCube~\cite{Abbasi:2011ss}
and LVD~\cite{Bari:1989nw}. 
In this paper we will consider the following laboratory sites: Kamioka
(SK and KamLAND), the South Pole (IceCube), Sudbury
(SNO+)~\cite{Kraus:2010zzb} and ANDES\cite{Andes-Website}. In Table
\ref{table:detector-locations} we show the positions of the detectors
we consider in this work.
%

\begin{table}[!h]
\begin{centering}
\begin{tabular}{|c|c|c|c|}
\hline 
Site & Latitude & Longitude & Shadowing Prob. \\
     &          &           & Mantle \ (Core)  \tabularnewline
\hline 
Kamioka, Japan & 36.42$^o$N & 137.3$^o$ E &  0.559 (0.103) \tabularnewline
South Pole & 90$^o$N & - &  0.413  (0.065)\tabularnewline
ANDES & 30.25$^o$S & 68.88$^o$W & 0.449  (0.067)\tabularnewline
SNO, Canada  & 46.476$^o$N & 81.20$^o$E  & 0.571 (0.110)\tabularnewline
\hline
\end{tabular}
\par\end{centering}
\caption{Positions of the detectors we consider in this paper.
In the last column, we show the shadowing probability, 
the probability that SN neutrinos will pass only through the mantle 
(indicated as Mantle)  or will pass both through the mantle and core 
(indicated as Core).}
\label{table:detector-locations}
\end{table}

Following Ref.~\cite{Mirizzi:2006xx}, we have calculated  
the {\it shadowing} 
probabilities for the cases where $N$ ($N=1,2,3,4$) detectors are
considered simultaneously.
The shadowing probability is defined as the 
probability that a given detector or combination of 
detectors receives neutrinos from a galactic SN 
with Earth matter effect either by passing 
only through the mantle or both through the core and the mantle 
of the Earth~\cite{Mirizzi:2006xx}. 

Using the same model of SN distribution in the Milky Way considered in
Ref.~\cite{Mirizzi:2006xx}, which is based on the neutron star
distribution (reproduced in Eqs.~(\ref{eq:SN-distribution1}) and
(\ref{eq:SN-distribution2}) of the Appendix \ref{appendix-shadowing})
we can compute the shadowing probability for an arbitrary number of
detector positions on the Earth.

The shadowing probability for the most simple case where only a 
single detector is considered is shown in the last column of 
Table \ref{table:detector-locations}.
For the case where two detector locations, Kamioka and South Pole, 
are considered simultaneously, we show the results 
in Table~\ref{table:table-shadowing-2-detectors}. 
We observe that the numbers shown in this table agree well with 
the ones found in Table 2 of Ref.~\cite{Mirizzi:2006xx}. 
From Table~\ref{table:table-shadowing-2-detectors} we 
conclude that the probability that only one of these detectors observes  
SN neutrinos having passed through the Earth is about 72\%.  

\begin{table}[!h]
\begin{centering}
\begin{tabular}{|c|c|c|c|}
\hline 
 & \multicolumn{2}{c|}
{Earth Matter Effect} &\tabularnewline \hline 
 Case &\ Kamioka \ & South Pole\ \ & Shadowing Prob. \\
      &            &            & Mantle \ (Core)  \tabularnewline
\hline 
(1) & No & No &  0.152 (0.832) \tabularnewline
(2) & Yes & No & 0.435 (0.104) \tabularnewline
(3) & No & Yes & 0.288 (0.065) \tabularnewline
(4) & Yes & Yes & 0.125 (0.000)  \tabularnewline
\hline
\end{tabular}
\par\end{centering}
\caption{Earth shadowing probability for the case where two 
detectors at Kamioka and South Pole are considered. }
\label{table:table-shadowing-2-detectors}
\end{table}

\begin{table}[!h]
\vglue 0.1cm
\begin{centering}
\begin{tabular}{|c|c|c|c|c|}
\hline 
 & \multicolumn{3}{c|}
{Earth Matter Effect}&\tabularnewline
\hline 
Case &\ Kamioka \ & South Pole &\  ANDES\ \  & Shadowing Prob. \\
      &           &            &  & Mantle (Core) \tabularnewline
\hline 
(1) & No & No & No & 0.024 (0.767) \tabularnewline
\hline 
(2) & Yes & No & No & 0.388 (0.105)\tabularnewline
(3) & No & Yes & No & 0.034 (0.061)\tabularnewline
(4) & No & No & Yes & 0.128 (0.063)\tabularnewline
\hline 
(5) & Yes & Yes & No & 0.106 (0.000)\tabularnewline
(6) & No & Yes & Yes & 0.254 (0.003)\tabularnewline
(7) & Yes & No & Yes & 0.047 (0.000)\tabularnewline
\hline 
(8) & Yes & Yes & Yes & 0.020 (0.000)\tabularnewline
\hline
\end{tabular}
\par\end{centering}
\caption{Earth shadowing probability for the case where detectors at
  Kamioka, South Pole and Andes sites are considered.  
}
\label{table:table-shadowing-3-detectors-mantle}
\end{table}

\begin{table*}[!t]
\begin{centering}
\begin{tabular}{|c|c|c|c|c|c|}
\hline 
 & \multicolumn{4}{c|}
{Earth Matter Effect}&\tabularnewline
\hline 
Case &\ Kamioka \ & South Pole &\ \  ANDES\ \  &\ \ Sudbury \ \ & Shadowing Prob. \\
     &            &            &       &     & Mantle (Core)  \tabularnewline
\hline 
(1) & No & No & No & No & 0.008 (0.657) \tabularnewline
\hline 
(2) & Yes & No & No & No & 0.206 (0.105) \tabularnewline
(3) & No & Yes & No & No & 0.034 (0.061) \tabularnewline
(4) & No & No & Yes& No & 0.001 (0.063) \tabularnewline
(5) & No & No & No& Yes & 0.016 (0.111) \tabularnewline
\hline 
(6) & Yes & Yes & No & No & 0.205 (0.000) \tabularnewline
(7) & Yes & No  & Yes & No & 0.000 (0.000) \tabularnewline
(8) & Yes & No  & No & Yes &  0.282 (0.000) \tabularnewline
(9) & No & Yes  & Yes & No &  0.163  (0.003) \tabularnewline
(10) & No & Yes  & No & Yes &  0.000 (0.000) \tabularnewline
(11) & No & No  & Yes & Yes &  0.127 (0.000) \tabularnewline
\hline 
(12) & No & Yes  & Yes & Yes &  0.091 (0.000) \tabularnewline
(13) & Yes & No  & Yes & Yes &  0.047 (0.000) \tabularnewline
(14) & Yes & Yes  & No & Yes &  0.011 (0.000) \tabularnewline
(15) & Yes & Yes  & Yes & No &  0.012 (0.000) \tabularnewline
\hline 
(16) & Yes & Yes  & Yes & Yes &  0.008 (0.000) \tabularnewline
\hline
\end{tabular}
\par\end{centering}
\caption{Earth shadowing probability for the case where detectors at
  Kamioka, South Pole, ANDES and Sudbury sites are considered.}
\label{table:table-shadowing-4-detectors-mantle}
\end{table*}

Next in Table~\ref{table:table-shadowing-3-detectors-mantle} we show
the Earth shadowing probabilities for the case where we consider three
detectors: at Kamioka, South Pole and ANDES sites.
From this table, we see that the probability of having at least one of 
these detectors  observing SN neutrinos passing through the Earth while 
at least one of the other two sees them non shadowed by the Earth is 96\%, 
which is about 30\% larger than the case above with two detectors, one at 
Kamioka and the other at the South Pole. 
One can also compare our three detector combination with any two
detector combination found in Table 2 of Ref.~\cite{Mirizzi:2006xx}
where the largest probability for having one detector shadowed and 
one non shadowed is 87.2\%, occurring for the Pyh\"asalmi and South Pole sites.

In Table \ref{table:table-shadowing-4-detectors-mantle} we show the
case where four detectors at Kamioka, South Pole, ANDES and Sudbury sites
are considered.  With four detectors, the probability that at least
one of the detectors have the Earth effect and at least one of the
others have not, increases to 98\%.
We also found that the probability that at least one 
of these sites receives a SN neutrino flux which 
passes the core of the Earth is not very small, $\sim$ 34\%.

\subsection{Quantifying the Earth matter effect: 
Comparing the detectors with and without Earth matter effect}

Let us now try to quantify the Earth matter effect which could be
observed in a model independent way by comparing the yields of two (or
more) detectors if only some (not all) of them receive SN neutrinos
passing through the Earth's interior.  See
~\cite{Dighe:1999bi,Lunardini:2001pb,Dighe:2003jg,Dighe:2003vm} for
the previous works.

In this paper, we focus on the possible Earth matter effect for
$\bar{\nu}_e$ due to the larger number of expected events.  Since we
know now that $\theta_{13}$ is not too small, $\sin^2 2 \theta_{13}
\simeq 0.09-0.1$~\cite{Abe:2011sj,Adamson:2011qu,
  Abe:2011fz,An:2012eh,collaboration:2012nd}, it is expected that
Earth matter effect can only be large for the normal mass hierarchy
in the standard scenario we consider in this work~\cite{Dighe:1999bi}.
So we will assume here only the normal hierarchy in such a
standard scenario.
Note that for the case of $\bar \nu_e$ in the normal mass hierarchy,
collective effects, shock wave, etc. are in general expected to be small.

If SN neutrinos reach the detector after passing through the Earth
matter, assuming the impact of non-zero $\theta_{13}$ is not very
significant for the Earth matter effect itself (though the impact of
$\theta_{13}$ is expected to be large for the oscillations inside the SN,
affecting significantly $\bar{p}(E)$ especially for the 
case of the inverted mass hierarchy), the SN flux
spectrum is expected to get modified as follows~\cite{Dighe:1999bi},
\begin{equation}
F_{\bar{\nu}_e}^\oplus (E)
= \bar{p}^\oplus(E) F_{\bar{\nu}_e}^0(E)
+ [1-\bar{p}^\oplus(E)] F_{\bar{\nu}_x}^0(E),
\label{eq:SN-flux-earth}
\end{equation}
where 
\begin{widetext}
\begin{equation}
\bar{p}^\oplus (E)
= \frac{1}{ |U_{e2}|^2-|U_{e1}|^2 }
\left[ 
\left\{ |U_{e2}|^2-\bar{p}(E) \right\}\bar{p}^\oplus_{1e}
+ \left\{ \bar{p}(E) - |U_{e1}|^2\right\} \bar{p}^\oplus_{2e}
\right],
\end{equation}
\end{widetext}
where $U_{ek}$ ($k=1,2$) are the elements of neutrino mixing matrix 
which relate flavor and mass eigenstates
and 
\begin{equation}
\bar{p}^\oplus_{ke} \equiv P^\oplus(\bar{\nu}_k \to \bar{\nu}_e, L), \  \ 
(k = 1,2), 
\label{eq:Pke}
\end{equation}
are the probabilities that a mass eigenstate $\bar{\nu}_k$ $(k=1,2)$ 
entering the Earth will be detected as $\bar{\nu}_e$ 
at the detector, after traveling the distance $L$ inside the Earth. 

\begin{figure*}[!t]
\begin{center}
\vglue -1.0cm
\hglue -0.4cm
\includegraphics[width=0.98\textwidth]{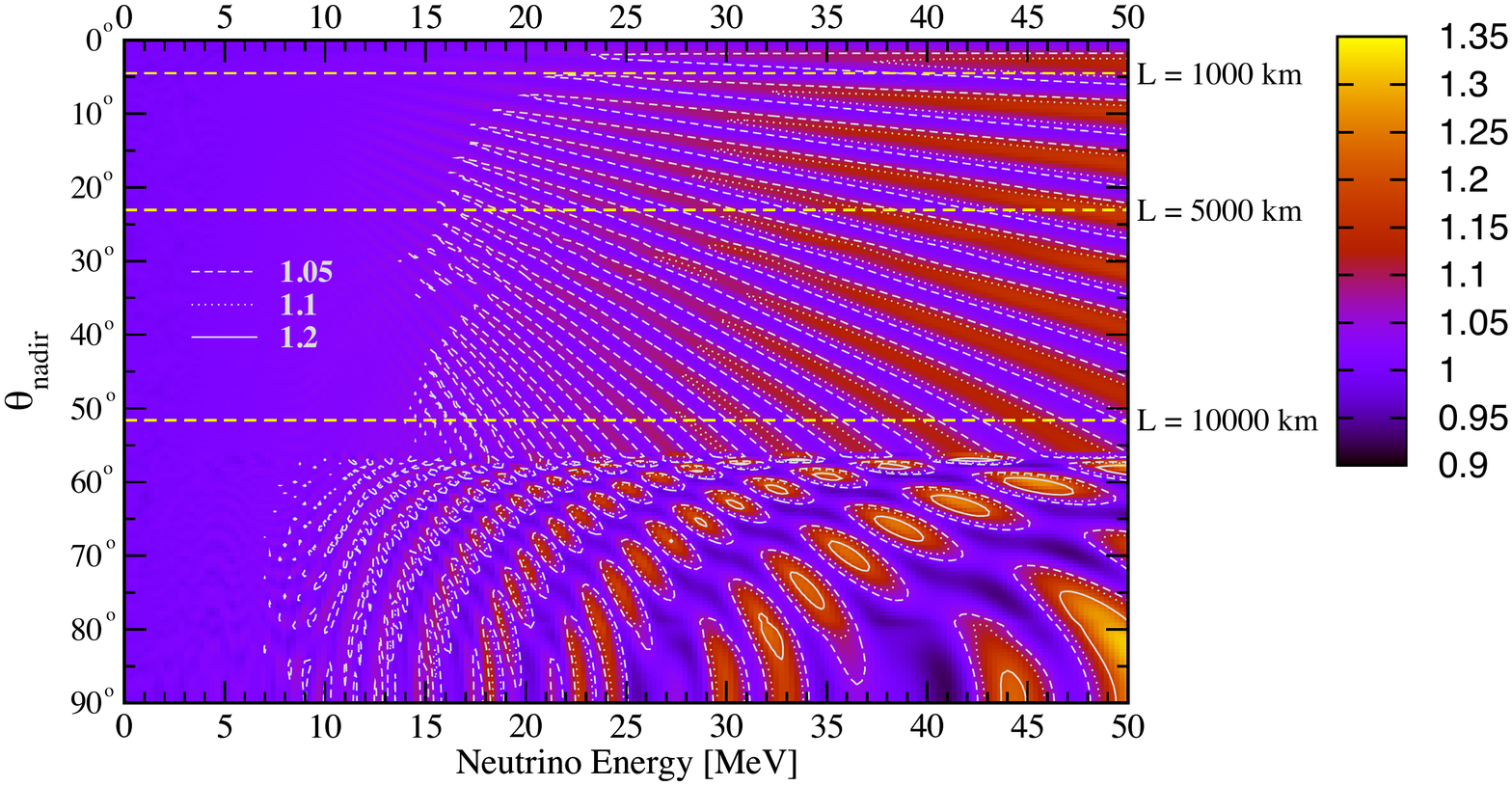}
\end{center}                          
\vglue -2.5cm
\caption{Neutrino Oscillogram, iso-contour plot of the ratio 
$P^\oplus(\bar{\nu}_1 \to \bar{\nu}_e)/c_{12}^2$ in the plane of 
the neutrino energy and the nadir angle.
Note that unity for this ratio corresponds to the case where the Earth
matter effect is absent.  }
\label{fig:SN-oscillogram}
\end{figure*}

If we take the difference of SN spectra given in 
Eqs. (\ref{eq:SN-flux-earth}) and (\ref{eq:SN-flux-vac}) 
with and without Earth matter effect, 
\begin{widetext}
\vglue -0.8cm
\begin{eqnarray}
\Delta F_{\bar{\nu}_e} 
& \equiv & 
F_{\bar{\nu}_e}^\oplus (E) - F_{\bar{\nu}_e}(E) \nonumber \\
& = & \frac{1}{ |U_{e1}|^2-|U_{e2}|^2 } 
\left\{ [2\bar{p}(E) - 1] \left(|U_{e2}|^2- \bar{p}^\oplus_{2e}\right)
+|U_{e3}|^2 \left( \bar{p}(E)-\bar{p}^\oplus_{2e} \right) \right\} 
\left\{F_{\bar{\nu}_e}^0(E) - F_{\bar{\nu}_x}^0(E)\right\}  
\nonumber \\
& \simeq &\frac{1}{\cos 2\theta_{12}} 
[ 2 \bar{p}(E) - 1 ] (s^2_{12} - \bar{p}^\oplus_{2e}) 
\left\{F_{\bar{\nu}_e}^0(E) - F_{\bar{\nu}_x}^0(E)\right\} \nonumber \\
& \simeq &
(\bar{p}^\oplus_{1e} - c^2_{12}) 
\left\{F_{\bar{\nu}_e}^0(E) - F_{\bar{\nu}_x}^0(E)\right\},
\label{eq:Delta_F}
\end{eqnarray}
\end{widetext}
where $\bar{p}(E) \approx c_{12}^2$ was assumed to get 
the last expression. We also observe  that the term 
proportional to $\vert U_{e3}\vert^2$ can be dropped because 
it will only contribute to about 7\%  of the first term.
As one can see from Eq. (\ref{eq:Delta_F}), in order to observe the
Earth matter effect, the deviation of $\bar{p}^\oplus_{1e}$ from
$c^2_{12}$ must be large enough and at the same time, the difference
between the original spectra of $F_{\bar{\nu}_e}^0(E)$ and
$F_{\bar{\nu}_x}^0(E)$ must be also significantly large.

In order to have some idea about the magnitude of the Earth matter
effect as functions of the neutrino energy and the incident angle of
SN neutrinos, we show in Fig.~\ref{fig:SN-oscillogram} the
iso-contours of the quantity $P^\oplus(\bar{\nu}_1 \to
\bar{\nu}_e)/c_{12}^2$ in the plane of the nadir angle,
$\theta_{\text{nadir}}$, and the neutrino energy where
$P^\oplus(\bar{\nu}_1 \to \bar{\nu}_e)$ was obtained by numerically
integrating the neutrino evolution equation using the matter density
profile of the Earth based on PREM (Preliminary Reference Earth
Model)~\cite{Dziewonski:1981xy}.  Note that any deviation of
$P^\oplus(\bar{\nu}_1 \to \bar{\nu}_e)/c_{12}^2$ from unity implies
the presence of the Earth matter effect.

\begin{figure}[!t]
\vglue -0.9cm
\begin{center}
\hglue -0.3cm
\includegraphics[width=0.70\textwidth]{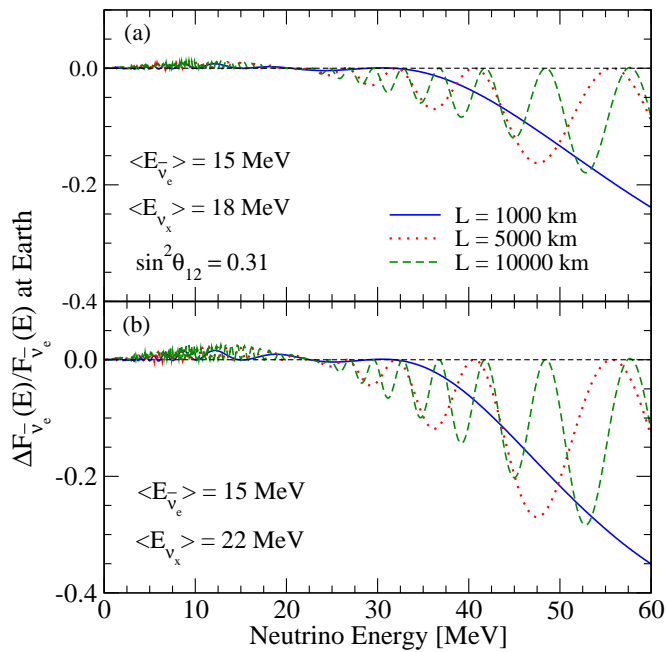}
\end{center}                          
\vglue -0.8cm
\caption{The fractional difference of the expected neutrino flux 
spectra with and without the Earth matter effect, 
$\Delta F_{\bar{\nu}_e}/F_{\bar{\nu}_e}$ where 
$\Delta F_{\bar{\nu}_e}  \equiv 
F_{\bar{\nu}_e}^\oplus (E) - F_{\bar{\nu}_e}(E)$,  
given in Eq. (\ref{eq:Delta_F}). 
The SN neutrino parameters are the same as assumed 
in Fig.~\ref{fig:SN-spec-10kpc}. 
}
\label{fig:SN-spec-frac-diff-10kpc}
\end{figure}

This plot is similar to the so called ``neutrino oscillogram'' studied in 
detail in Ref.~\cite{nu-oscillogram} in the context of atmospheric neutrinos.
As we can see from this plot, the Earth effect is as expected 
strongest when neutrinos pass through the Earth's core (corresponding to
$\theta_{\text{nadir}} \lsim 33^\circ$)
and for higher neutrino energies.
Here $\theta_{\text{nadir}}$ is defined such that 
$\theta_{\text{nadir}} = 0^\circ$
corresponds to the case 
where SN neutrinos arrive at the detector from the other side 
of the Earth passing the 
center of the Earth and 
$\theta_{\text{nadir}} = 90^\circ$ corresponds to %
the case where neutrinos come from the horizontal direction.

In Fig.~\ref{fig:SN-spec-frac-diff-10kpc} we show the fractional
difference of the flux spectra with and without the Earth effect
defined by $\Delta F_{\bar{\nu}_e}/F_{\bar{\nu}_e}$ where
$\Delta F_{\bar{\nu}_e}$ is given by Eq. (\ref{eq:Delta_F}) for 
for three different path-lengths 
in the Earth, $L= 1000$ km (solid blue curve), 
$L= 5000$ km (dotted red curve) and 
$L= 10000$ km (dashed green curve).
From this figure, we can see that strong Earth matter
effect is expected in the higher energy range.  We must note, however,
that as the energy becomes higher, the number of events gets smaller, so
that in order to identify the Earth matter effect, both the Earth
effect (difference of probabilities with and without matter effect)
and the number of events in the relevant energy range must be large
enough.

\begin{table*}[!t]
\begin{centering}
\begin{tabular}{|c|c|c|c|c|c|c|}
\hline 
& {$~E < 30$ MeV~} &  $~30<E/{\rm MeV} <40~$ & $~40<E/{\rm MeV} < 50~$ &   $~E > 50$ MeV~ & 
Case & Compatibility   \\ 
\hline \hline
Vacuum (observed) & 18159 $\pm$ 135 & 4973 $\pm$ 71 & 2032 $\pm$ 45 &
889 $\pm$ 30 & $\langle E_{\nu_x} \rangle $ = 22 MeV & $3.1 \sigma$
\\ 
1000 km (prediction)& 18132 $\pm$ 374 & 5065 $\pm$ 198 & 1908 $\pm$
121 & 700 $\pm$ 74& $\beta_x=\beta_e=4$ & \\ \hline 
Vacuum (observed) & 17395 $\pm$ 132 & 5785 $\pm$ 76 & 2583 $\pm$ 51 & 
1147 $\pm$ 34 & $\langle E_{\nu_x} \rangle $ = 22 MeV &
$2.2 \sigma$\\ 
1000 km (prediction) & 17370 $\pm$ 367 & 5858 $\pm$ 213 & 2483 $\pm$ 139 & 
988 $\pm$ 87 & $\beta_x=4$ $\beta_e=3$ & \\ \hline 
Vacuum (observed) & 16031 $\pm$ 127 & 6674 $\pm$ 82 & 3594 $\pm$ 60 & 
1978 $\pm$ 45 & $\langle E_{\nu_x} \rangle $ = 22 MeV & $0.7
\sigma$\\ 
1000 km (prediction) & 16011 $\pm$ 352 & 6728  $\pm$ 228 & 3541 $\pm$ 166 & 
1917 $\pm$  122 & $\beta_x=4$ $\beta_e=2$ & \\ \hline 
Vacuum (observed) & 16863 $\pm$ 130 & 5722 $\pm$ 76 & 2864 $\pm$ 54 & 
1604 $\pm$ 40  & $\langle E_{\nu_x} \rangle $ = 22 MeV & $3.2\sigma$\\ 
1000 km (prediction) & 16837 $\pm$ 361 & 5787 $\pm$ 212 & 2731 $\pm$ 145 & 
1321 $\pm$ 101 & $\beta_x=\beta_e=3$ & \\ \hline 
Vacuum (observed) & 15499 $\pm$ 125 & 6611 $\pm$ 81 & 3875 $\pm$ 62&
2434 $\pm$ 49 & $\langle E_{\nu_x} \rangle $ = 22 MeV & $1.7 \sigma$\\ 
1000 km (prediction) &15479 $\pm$ 346 & 6657 $\pm$ 227 &3789 $\pm$ 171 &
2250 $\pm$ 132 & $\beta_x=3$ $\beta_e=2$ & \\ \hline 
Vacuum (observed) & 14790$\pm$122 & 6388$\pm$ 80 & 4089$\pm$ 64& 
3059$\pm$ 55 & $\langle E_{\nu_x} \rangle $ = 22 MeV & $2.8 \sigma$\\ 
1000 km (prediction) & 14766$\pm$338 & 6419$\pm$223 & 3971$\pm$ 175& 
2701$\pm$ 145& $\beta_x=\beta_e=2$ &\\ \hline \hline 
Vacuum (observed) & 17686$\pm$133 & 5240$\pm$72 & 2439$\pm$49 & 
1285 $\pm$36 & $\langle E_{\nu_x} \rangle $ = 24 MeV & $4.3 \sigma$ \\ 
1000 km (prediction) & 17655$\pm$370 & 5343$\pm$203 & 2272$\pm$133 & 
990$\pm$88 & $\beta_x=\beta_e=4$ &\\ \hline 
Vacuum (observed) & 16922$\pm$130 & 6052$\pm$78 & 2990$\pm$55 & 
1543$\pm$39 & $\langle E_{\nu_x} \rangle $ = 24 MeV & $3.1 \sigma$ \\ 
1000 km (prediction) & 16892$\pm$362 & 6136$\pm$218 & 2847$\pm$ 148 & 
1278$\pm$100 & $\beta_x=4$ $\beta_e=3$ & \\ \hline
Vacuum (observed) & 15557$\pm$125 & 6941$\pm$83 & 4001$\pm$63 & 
2374$\pm$49 & $\langle E_{\nu_x}\rangle $ = 24 MeV & $1.7 \sigma$\\ 
1000 km (prediction) & 15533$\pm$347 & 7006$\pm$233 & 3905$\pm$174 & 
2207$\pm$ 131& $\beta_x=4$ $\beta_e=2$ & \\ \hline 
Vacuum(observed) & 16441$\pm$128 & 5858$\pm$77 & 3174$\pm$56 & 
2022$\pm$45 & $\langle E_{\nu_x} \rangle $ = 24 MeV & $4.0 \sigma $\\ 
1000 km (prediction) & 16409 $\pm$356 & 5928$\pm$214 & 3007$\pm$ 153&
1625$\pm$112 & $\beta_x=\beta_e=3$ & \\ \hline 
Vacuum (observed) & 15077$\pm$123 & 6746$\pm$82 & 4185$\pm$65 & 
2853$\pm$53 & $\langle E_{\nu_x} \rangle $ = 24 MeV & $2.5\sigma$ \\ 
1000 km (prediction) & 15051$\pm$341 & 6797$\pm$229 &4065$\pm$177 &
2554$\pm$141 & $\beta_x=3$ $\beta_e=2$ & \\ \hline 
Vacuum (observed) & 14439$\pm$120 & 6400$\pm$80 &4248$\pm$65 &
3402$\pm$58 & $\langle E_{\nu_x} \rangle $ = 24 MeV & $3.3\sigma$\\ 
1000 km (prediction) & 14410$\pm$334 &6430$\pm$223 &4116$\pm$179 & 
2948$\pm$151 & $\beta_x=\beta_e=2$ & \\ \hline
\end{tabular}
\par\end{centering}
\caption{Number of inverse $\beta$-decay events expected at the
  SK detector (1.7 $\times 10^{33}$ free protons) for a SN
  happening at 5 kpc from the Earth for $E< 30$ MeV, $30<E/{\rm
    MeV}<40$, $40<E/{\rm MeV}<50$ and $E > 50$ MeV for the case of
  vacuum and matter effect with a baseline of 1000 km and various SN
  parameters.  We assume $L_{{\bar{\nu_e}}}/L_{\nu_x} = 1$ and
  $\langle E_{\bar \nu_e}\rangle$ = 15 MeV.  We assume that the SK
  detector receives SN neutrinos without the Earth matter effect
  whereas the ANDES neutrino detector receives them after traveling
  1000 km inside the Earth.  The numbers in the row indicated as
  vacuum are the ones to be observed at SK detector which must be
  compared with the theoretical prediction at SK  inferred from the
  observed number of events at ANDES neutrino detector.
In the last column we point out in how many $\sigma$ the 1000 km
observation is distinguishable from vacuum. }
\label{table:SN-SK-events}
\end{table*}

Since we cannot compare the number of events at the same detector with
and without Earth matter effect, we need two or more detectors to be
able to conclude something on the matter effect.  For simplicity and
for the sake of discussion, let us consider only two detectors at two
different sites, say, one at Kamioka (SK) and other at ANDES.
 
Suppose that the arrival of the SN neutrinos at the ANDES
  detector is shadowed by the Earth, while at SK detector at Kamioka
  is not.
Then, to some extent, within the statistical and
systematic errors, one can try to infer the expected SN spectra at SK
from the observed ones at ANDES (or vice versa) as if the SK detector
{\it were also shadowed} like ANDES.  If both detectors receives SN
neutrinos without Earth matter effect these two spectra must agree
with each other but with the matter effect, they are not expected to
coincide exactly.

In order to see the presence of Earth matter effect, 
the combination of SK and ANDES must be able to distinguish, for a
given SN model, vacuum from matter event distribution. To illustrate
that, we show in Table~\ref{table:SN-SK-events} the expected number of
events in SK for four different energy bins for vacuum and matter with
$L= 1000$ km, for a variety of SN parameters.  If we assume that the
shadowed SN neutrino spectra distribution of events at SK can be
provided by ANDES with an uncertainty in each bin equal to the
statistical uncertainty in ANDES we can estimate in which of these
cases one can establish the presence of matter effects.  In fact, since
the first bin $E<$ 30 MeV is not sensitive to matter effects, one can
use it as a control bin to evaluate the accuracy of the vacuum
distribution provided by ANDES data.  For this study, we consider SN
event at 5 kpc from the Earth as we need larger number of events for
ANDES.

To give a quantitative idea of the discrimination power, we have
assumed, for each SN model in Table~\ref{table:SN-SK-events} the vacuum
distribution to be the hypothesis to be tested with data at 1000
km. We use the last two energy bins in order to estimate the deviation
from vacuum in terms of standard deviations.  These numbers are given
in the last column of Table~\ref{table:SN-SK-events}.  For most cases,
matter effects can be established in more than 2 $\sigma$.  We have
also verified that if $\langle E_{\nu_x}\rangle \lsim 18$ MeV we
cannot distinguish matter effects for any value of the other SN
parameters considered in Table~\ref{table:SN-SK-events}, at least if
$L_{{\bar{\nu_e}}}/L_{\nu_x} = 1$.

\section{Discussions and Conclusions}
\label{sec:conclusions}

The ANDES laboratory, if constructed, will be the first deep
underground laboratory in the Southern Hemisphere. 
It can offer the possibility to explore rare physics events 
that can profit from the low natural background environment
because of the overburden of $\sim 4.5$ kmwe.
In particular, dedicated neutrino
physics and dark matter search programs, that could also benefit from
the its unique geographic location, could be envisaged.  See
\cite{Andes-Website} for updates on the status of the laboratory.

In this work, we have studied the potential of a few kt liquid scintillator 
neutrino detector at the ANDES underground laboratory for neutrino 
astrophysics and geophysics. 
Since there are very few nuclear reactors in South America, 
the location of the ANDES laboratory is specially suitable for 
geoneutrino observations. 
Moreover, due to thick continental crust around the laboratory, 
higher geoneutrino event rate is expected,  
substantially larger than at Kamioka and Gran Sasso, 
which by itself would be interesting to confirm experimentally.

Concerning the observations of neutrinos coming from SN, the ANDES
neutrino detector could play an important role.
First of all, because of the small event rate of the nearby 
SN (within a few 10 kpc from the Earth or so), 
having as many large neutrino detectors as possible is  highly desirable.  
Furthermore, because of the location, the presence of the ANDES detector 
will increase the chance of observing the Earth matter effect by combining 
the signal at the ANDES with other detectors in the Northern Hemisphere.
The ANDES neutrino detector could also integrate the international
SN watch network SNEWS~\cite{Antonioli:2004zb}. 

We have focused here on galactic SN neutrinos and geoneutrinos.
However, with such a liquid scintillator detector one could also try
to study solar neutrinos in a similar fashion as 
Borexino~\cite{Borexino-solar} 
and KamLAND~\cite{Abe:2011em} 
have done and as SNO+~\cite{Kraus:2010zzb} 
 plans to do. In fact, Borexino has shown how
pep, CNO and perhaps even pp solar neutrinos can be accessible by such
a detector.  There is also an interesting proposal to search for
indirect dark matter signals through neutrinos which are coming from
the dark matter annihilation in the Sun in these type of detectors
(see e.g.~\cite{Kumar:2011hi}).

\begin{acknowledgments} 
  \vspace{-0.3cm} This work is supported by Funda\c{c}\~ao de Amparo
  \`a Pesquisa do Estado de S\~ao Paulo (FAPESP), Funda\c{c}\~ao de
  Amparo \`a Pesquisa do Estado do Rio de Janeiro (FAPERJ), by
  Conselho Nacional de Ci\^encia e Tecnologia (CNPq) and by the European 
Commission under the contract PITN-GA-2009-237920.  
We thank Xavier Bertou for useful correspondence on 
the current status of the ANDES laboratory.
H.N. and R.Z.F. thank  the Galileo Galilei Institute for Theoretical Physics
 for the hospitality and  Belen Gavela and Silvia Pascoli 
for the invitation to the GGI workshop: What is $\nu$? 
at the Galileo Galilei Institute for Theoretical Physics 
and the INFN for partial support during the completion of this work. 
R.Z.F. also acknowledges partial support from the European Union FP7 ITN 
INVISIBLES (PITN-GA-2011-289442).
\end{acknowledgments}

\appendix 

\section{Calculations of the number of event induced by
the inverse beta decay reaction}
\label{appendix-inverse-beta}

The number of event induced by the inverse beta decay reaction 
$\bar{\nu}_e + p \to n + e^+$, 
is given by 
\begin{equation}
N
= N_p
\int_{E_{\text{min}}}^\infty dE\  F_{\bar{\nu}_e}(E)
\sigma_{\bar{\nu}_e p} (E)
\end{equation}
where $E_{\text{min}} = 1.806$ MeV
is the threshold of this reaction,
$N_p$ is the number of free protons in the detector and 
$E$ is the observed energy.
For simplicity, we assume perfect energy resolution, which is
a good approximation for the results presented in this work. 
While the accurate absorption cross section of $\bar{\nu}_e$ on 
proton, $\sigma_{\bar{\nu}_e p} (E)$, can be found in 
Ref.~\cite{Strumia:2003zx}, we only quote the simple 
approximate analytic expression given in this reference, 
which is sufficient for our purpose, 
\begin{eqnarray}
\sigma_{\bar{\nu}_e p} (E) & \approx & p_e E_e 
E^{-0.07056 + 0.02018 \ln E - 0.001953 \ln^3 E}
\nonumber \\
& &  \times 10^{-43} 
\ \ [\text{cm}^2], 
\end{eqnarray}
where $E_e = E - (m_n-m_p) \simeq E - 1.293 $ MeV ($m_n$, $m_p$ are 
mass of neutron and proton, respectively), and $p_e$ is the 
momentum of positron, all the energies should be given in MeV.

\section{Calculations for proton-neutrino scattering}
\label{appendix-proton}

The differential cross section for $\nu + p \to \nu + p$, 
for any neutrino flavor, is  given by~\cite{Beacom:2002hs,Weinberg:1972tu},
\begin{equation}
\frac{d \sigma}{dT} 
= \frac{G_F^2 m_p}{\pi}
\left[ \left(1 - \frac{m_p T}{2E^2} \right) c_v^2 
+ 
 \left(1 + \frac{m_p T}{2E^2} \right) c_a^2 
\right],
\end{equation}
where $G_F$ is the Fermi constant, $m_p$ is the proton mass,  
$T$ is the kinetic energy of recoil proton, 
$E$ is the neutrino energy, 
$c_v = 0.04$ and $c_a = 1.27/2$. 
To take into account the loss of proton energy, we calculate
the quenched proton energy, $T'$, by 
\begin{equation}
T'(T) 
= \int_0^T \frac{dT}{1+k_B \langle dT/dx \rangle },
\end{equation}
where $k_B$ is called Birks constant. 
The numerical values of $\langle dT/dx \rangle$ which describe 
the energy loss of proton in scintillator were taken 
from Ref.~\cite{NIST}.

The event number distribution $dN/dT'$ is calculated 
by 
\begin{equation}
\frac{dN}{dT'} 
= N_p \left(\frac{dT'}{dT} \right)^{-1} 
\int_{E_{\text{min}}}^\infty dE \frac{dF}{dE}
\frac{d \sigma}{dT}\,,
\label{eq:dNdT'}
\end{equation}
where $E_{\text{min}}$ is the minimum energy of neutrino which 
can produce recoil proton with the kinetic energy $T$, given by 
\begin{equation}
E_{\text{min}} = \frac{1}{2} 
\left[T + \sqrt{ T(T+2m_p)} \right]
\simeq \sqrt{\frac{m_p T}{2}}.
\end{equation}
We further taken into account the detector energy resolution 
to convert $dN/dT'$ in Eq.~(\ref{eq:dNdT'}) into 
the ones actually observed, $dN/dT_{\text{que}}$ as 
\begin{equation}
\frac{dN}{dT_{\text{que}}}
= \int_0^\infty dT' \frac{dN}{dT'} 
R(T_{\text{que}},T')
\label{eq:dNdTobs},
\end{equation}
where $R(T_{\text{que}},T')$ is the resolution function 
given by
\begin{equation}
R(T_{\text{que}},T') 
= \frac{1}{\sqrt{2\pi} \sigma }
\exp \left[ - \frac{(T_{\text{que}}-T')^2}{2\sigma^2} \right]
\label{eq:resolution},
\end{equation}
with the resolution assumed to be 
5 \%/$\sqrt{T'/\text{MeV}}$ ~\cite{Dasgupta:2011wg}.

\section{Some details of the calculations of shadowing probabilities}
\label{appendix-shadowing}

For definiteness, we use the same galactic supernova distribution 
model considered in Eqs. (1) and (2) of Ref.~\cite{Mirizzi:2006xx}, 
which is based on the expected distribution of 
neutron stars in the Galaxy~\cite{Yusifov-Kucuk:2004} given by 
\begin{equation}
\sigma_{\text{SN}}(r) \propto r^4 \exp\left[-\frac{r}{1.25\ \text{kpc}} \right],
\label{eq:SN-distribution1}
\end{equation}
where $\sigma_{\text{SN}}(r)$ is the surface density of the core-collapse SN 
events as a function of the radial distance from the galactic center. 
We also take the same vertical distribution of the SN events as 
\begin{eqnarray}
R_{\text{SN}}(z) &\propto & 
0.79 \exp\left[-\left(\frac{z}{212\ \text{pc}}\right)^2 \right]  \nonumber \\
& & + 0.21 \exp\left[-\left(\frac{z}{636\ \text{pc}} \right)^2 \right],
\label{eq:SN-distribution2}
\end{eqnarray}
where $z$ is the vertical distance from the galactic plane so that the 
SN distribution $n_{\text{SN}}(r,z) \propto \sigma_{\text{SN}}(r) R_{\text{SN}}(z)$. 
%
\begin{figure}[!t]
\vglue -2.0cm
\hglue -0.2cm
\includegraphics[width=0.66\textwidth]{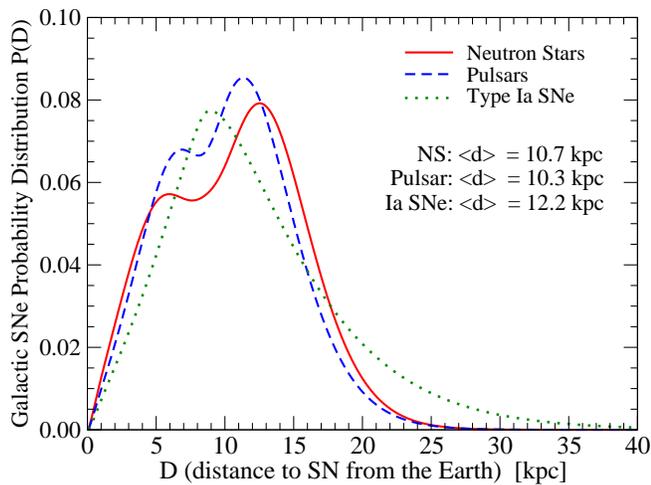}
\vglue -0.8cm
\caption{Probability of SN distribution in the Milky Way based on 
the distributions considered in Ref.~\cite{Mirizzi:2006xx}. 
The red solid, dashed blue and dotted green curves 
indicate, respectively, neutron star, pulsar and 
type Ia SN. In this work we consider the neutron star 
distribution (red curve) as our reference SN distribution. 
}
\label{prob-SN-distance}
\end{figure}

In Fig.~\ref{prob-SN-distance} we show the expected SN probability distribution 
in the Milky Way based on these distributions by the solid red curve. 
For comparison, we show the other distribution considered in 
Ref.~\cite{Mirizzi:2006xx} which are based on the pulsar (blue 
dashed curve) and type Ia SN (green dotted curve). 
We note that in calculating the shadowing probabilities 
rotation of the Earth was taken into account in the same way as 
in ~\cite{Mirizzi:2006xx}.

\end{document}